\documentclass[12pt]{book}
\usepackage[sectionbib]{natbib}
\usepackage{array,epsfig,fancyhdr,rotating}
\usepackage[]{hyperref}  

\usepackage[margin=1in]{geometry}

\usepackage{amsmath}
\usepackage{amssymb}
\usepackage{amsfonts}
\usepackage{multirow}
\usepackage{amsthm}
\usepackage{color}
\usepackage{amsmath,latexsym}
\usepackage{mathtools}
\usepackage{bm}
\DeclareMathOperator*{\argmax}{arg\,max}

\newcommand{\bc}{{\bf c}}
\newcommand{\be}{{\bf e}}

\newtheorem*{theorem*}{Theorem}
\newtheorem{theorem}{Theorem}

\newtheorem{algorithm}{Algorithm}

\newtheorem{cor}{\sc Corollary}

\begin{document}
\renewcommand{\baselinestretch}{1.5}

\renewcommand{\thefootnote}{}
$\ $\par


\fontsize{10.95}{14pt plus.8pt minus .6pt}\selectfont
\vspace{0.8pc}
\centerline{\large\bf MODULARITY BASED COMMUNITY DETECTION}
\vspace{2pt}
\centerline{\large\bf IN HETEROGENEOUS NETWORKS}
\vspace{.4cm}
\centerline{Jingfei Zhang and Yuguo Chen}
\vspace{.4cm}
\centerline{\it University of Miami and University of Illinois at Urbana-Champaign}
\vspace{.55cm}
\fontsize{9}{11.5pt plus.8pt minus .6pt}\selectfont


\begin{quotation}
\noindent {\it Abstract:}
Heterogeneous networks are networks consisting of different types of nodes and multiple types of edges linking such nodes. 
While community detection has been extensively developed as a useful technique for analyzing networks that contain only one type of nodes, very few community detection techniques have been developed for heterogeneous networks. 
In this paper, we propose a modularity based community detection framework for heterogeneous networks. Unlike existing methods, the proposed approach has the flexibility to treat the number of communities as an unknown quantity. We describe a Louvain type maximization method for finding the community structure that maximizes the modularity function. Our simulation results show the advantages of the proposed method over existing methods. Moreover, the proposed modularity function is shown to be consistent under a heterogeneous stochastic blockmodel framework.
Analyses of the DBLP four-area dataset and a MovieLens dataset demonstrate the usefulness of the proposed method.\par

\vspace{9pt}
\par
\end{quotation}\par

\fontsize{12}{14pt plus.8pt minus .6pt}\selectfont
\setcounter{chapter}{1}
\setcounter{equation}{0}
\noindent {\bf 1. Introduction}
\vspace{3pt}

Network community detection has attracted much recent attention from various scientific communities, including statistics, physics, information technology, biology, social science and many others.
A real-world network often displays a high level of inhomogeneity in its edge distribution, with high edge density within certain groups of nodes and low edge density between these groups. 
This feature is often referred to as the \textit{community structure} (Fortunato, 2010). Community structures have been observed in networks in social science, biology, political science and so on. 
For example, in a gene regulation network, communities are groups of genes that function together in biological processes to carry out specific functions (Zhang and Cao, 2017).
Detecting communities in real-world networks can help us better understand the architecture of the network. Further, it allows us to investigate the property in individual communities, which may be different from the aggregated property from the network as a whole.

Many community detection techniques have been proposed in recent years. See Fortunato (2010) for a comprehensive review. 
One class of methods involve maximizing some partition quality function over all possible partitions of the network (Shi and Malik, 2000; Newman and Girvan, 2004; Newman, 2006; Rohe et al., 2011). 
Another class includes model based approaches that estimate community structures through fitting probabilistic models to the observed networks (Airoldi et al., 2008; Bickel and Chen, 2009; Jin, 2015).
In the second class of approaches, we need to know the number of communities a priori.

\begin{figure}[!htb]
 \centering
\includegraphics[scale=0.125, trim=0 20mm 0 0]{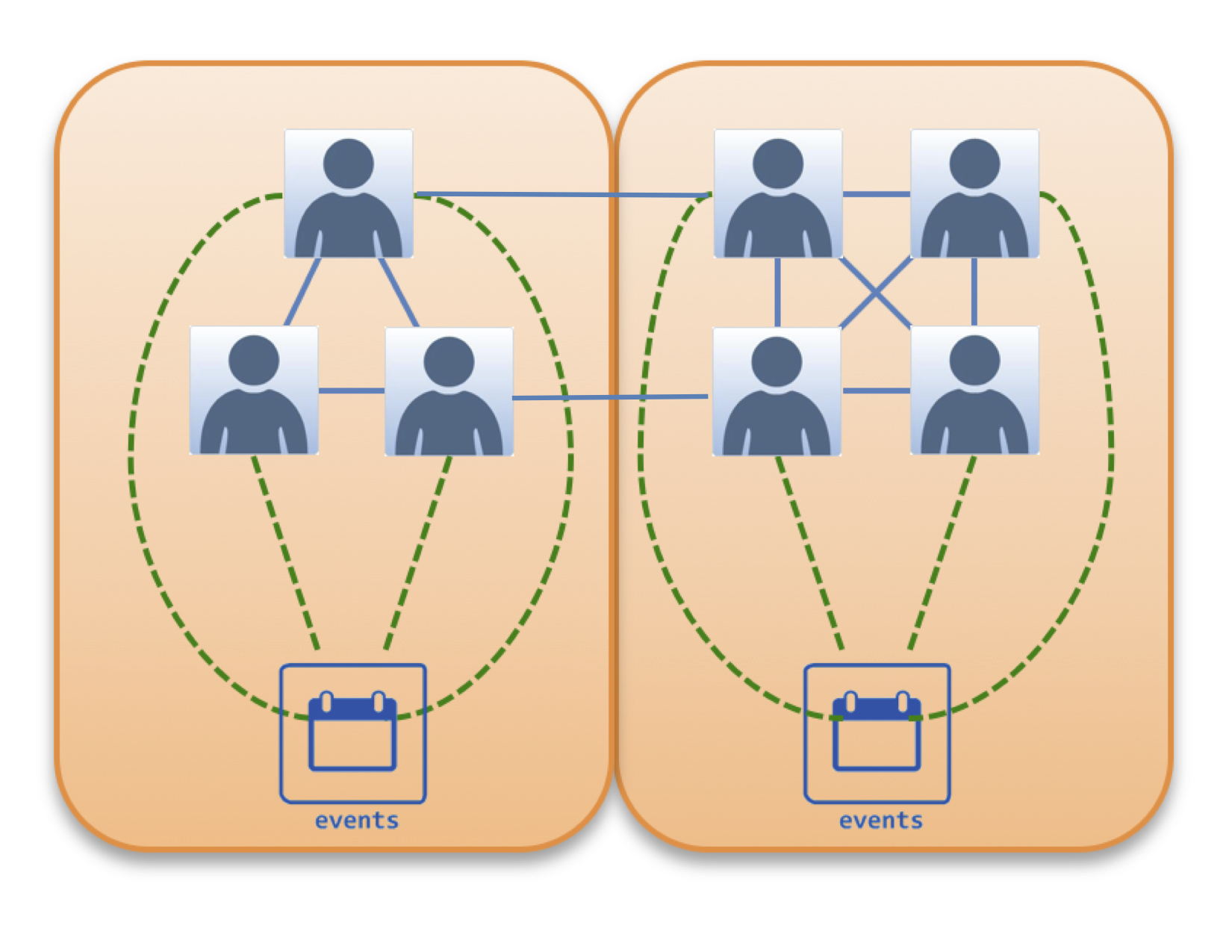}
\caption{Two communities in a heterogeneous Facebook network with two types of nodes: users and events. }
\label{comm}
\end{figure}

The existing community detection approaches primarily focus on homogeneous networks, i.e., networks with only one type of nodes. 
However, networks representing real-world complex systems often contain different types of nodes and different types of edges linking such nodes; we refer to such type of networks as \textit{heterogeneous networks}. 
For example, in a healthcare network, nodes can be patients, diseases, doctors and hospitals. 
The edges can be in the type of patient-disease (patient treated for disease), patient-doctor (patient treated by doctor), doctor-hospital (doctor works at hospital).
Figure~\ref{comm} provides a simple illustration of a heterogeneous network.
In this illustrative heterogeneous Facebook network, there are two types of nodes, users and events. Furthermore, there are two types of interactions in this network. A user is linked to another user through friendship and a user is linked to an event through attendance. 

To find communities in a heterogeneous network using the existing methods developed for homogeneous networks, there are two possible approaches. 
One approach is to treat the heterogeneous network as a homogeneous network. In this approach, we do not differentiate the different types of nodes and edges.
The other approach is to consider each type of nodes in the network separately, i.e., discard the information from the edges linking different types of nodes. 
In both approaches, we lose important information. 
In the first approach, we ignore the fact that different types of nodes may behave differently. 
For example, in Figure~\ref{comm}, users and events behave in different ways; a user can become friends with other users but an event can not link to other events. 
Using the first approach, the community detection algorithm does not distinguish the two different types of nodes. Losing such important information may lead to poor community detection results. 
In the second approach, the valuable information from edges that link different types of nodes are ignored. 
For example, in Figure~\ref{comm}, the user-event links show how different users are attracted to different events. 
Including such information can help us better identify the communities in users. 
Moreover, it provides important insights on the types of events that are clustered with each community of users.

To find community structures in heterogeneous networks, a preferable approach should take into account all the information contained in the heterogeneous network, including the different types of nodes, the homogeneous edges (edges that connect two nodes of the same type) and the heterogeneous edges (edges that connect two nodes of different types).
The objective of the approach is to cluster the nodes in the heterogeneous network into several non-overlapping groups such that there are more homogeneous and heterogeneous edges within these groups and fewer homogeneous and heterogeneous edges between these groups; see Figure~\ref{comm} for a simple illustration of a heterogeneous Facebook network with two communities. 

Several methods have been proposed recently for detecting communities in heterogeneous networks (Sun and Han, 2012; Liu et al., 2014; Sengupta and Chen, 2015). One limitation of the existing methods is that they may have requirements on the number of node types or edge types in the network (see for example Sun and Han (2012) and Liu et al. (2014)). Another limitation of the existing methods is that they may require the number of communities in the network to be pre-specified (see for example Sengupta and Chen (2015)). This requirement could be difficult to meet in practice, since we generally do not know the number of communities in real-world networks. 
Lastly, very large networks can be computationally challenging for some existing methods, such as the spectral clustering approach proposed in Sengupta and Chen (2015). 

In this paper, we propose a modularity based heterogeneous network community detection framework. 
Our contribution is threefold. First, we formally define a null model for a heterogeneous network. Under the proposed null model, we calculate the probabilities of having a homogeneous edge between two nodes of the same type and a heterogeneous edge between two nodes of different types. 
Second, we propose a Louvain type maximization method that can efficiently maximize the proposed modularity function. 
The application of the maximization method on a real-world network with about 20,000 nodes takes less than 20 seconds on a standard PC. 
Our proposed approach can be applied to heterogeneous networks of any type. Furthermore, the number of communities does not need to be specified and can be treated as an unknown quantity.
Third, we show that the proposed modularity function for heterogeneous networks is consistent under a heterogeneous stochastic blockmodel framework. The consistency properties of modularity functions formulated for bipartite or multipartite networks follow as special cases. This theoretical result fills an existing gap in the literature. 

The rest of this article is organized as follows. 
Section 2 introduces the null model for a heterogeneous network and the definition of a modularity function. 
Section 3 discusses the Louvain type modularity maximization technique. 
Section 4 shows the consistency of the modularity function under the a heterogeneous stochastic blockmodel framework.
Section 5 demonstrates the advantages of our proposed method through simulation studies.
Section 6 discusses the application of the proposed method on the DBLP four-area dataset and a MovieLens dataset.
Section 7 provides some concluding remarks and discussions.

\vspace{3pt}
\setcounter{chapter}{2}
\setcounter{equation}{0}
\noindent {\bf 2. Modularity Function for Heterogeneous Networks}
\vspace{3pt}

Let $\mathcal{G}=(\bigcup_{i=1}^LV^{[i]},\mathcal{E}\cup\mathcal{E}^+)$ denote a simple heterogeneous network (no self loops or multiple edges) with $L$ types of nodes.
Node set $V^{[i]}$ contains all the nodes of the $i$th type, $i=1,\ldots,L$. 
Edge set $\mathcal{E}$ denotes the set of edges between nodes of the same type and $\mathcal{E}^+$ denotes the set of edges between nodes of different types. 
A homogeneous network $G_i(V^{[i]},E^{[i]})$ can be formed within each node set $V^{[i]}$, where $E^{[i]}$ is the set of edges between nodes in $V^{[i]}$. 
By definition, we have $\mathcal{E}=\bigcup_{i=1}^LE^{[i]}$. 
When $\mathcal{E}=\emptyset$, the heterogeneous network $\mathcal{G}$ forms a multi-partite network, i.e., edges are only established between different types of nodes. 
In this paper, we use the terms \textit{network} and \textit{graph} interchangeably. 

Newman and Girvan (2004) defined a quality function, usually referred to as the \textit{modularity function}, for measuring the strength of a division of a homogeneous network into communities. 
Given a homogeneous network $G(V,E)$ with $n$ nodes, $m$ edges and a community assignment $\bm{e}=(e_1,\ldots,e_n)$, where $e_i\in\{1,\ldots,K\}$ is the community that node $i$ belongs to, the modularity function is defined as
\begin{equation}
Q(\bm{e},G)=\frac{1}{2m}\sum_{i,j}\left[A_{ij}-E(A_{ij})\right]\delta(e_i,e_j),
\label{modularity}
\end{equation}
where $\delta(r,s)=1$ if $r=s$ and 0 otherwise. 
Here $A_{ij}$ is the $(i,j)$th entry of the adjacency matrix $A$ of the network and the expectation $E(A_{ij})$ is calculated under some null model that describes networks with no community structure. It is easy to see that $Q(\bm{e},G)\in[-1,1]$.

The modularity function for homogeneous networks measures the difference between the observed number of intra-community edges and the expected number of intra-community edges under the null model. 
If the observed number of intra-community edges in the network is close to the expected value, the modularity $Q$ is close to 0. 
When $Q$ approaches 1, the observed number of intra-community edges is much higher than the expected value, and this indicates a strong community structure. 
Since the modularity function measures the ``strength" of community structure with respect to a network partition, the community membership of a network is identified by maximizing the modularity function $Q(\bm{e},G)$ with respect to $\bm{e}$. 
The number of communities $K$ does not need to be pre-specified in this approach and can be treated as an unknown quantity.

To introduce the modularity based community detection framework for heterogeneous networks, we focus on the case with only two types of nodes ($L=2$).
The framework can be easily generalized to networks that contain more than two types of nodes. 
For a heterogeneous network $\mathcal{G}=(V^{[1]}\cup V^{[2]}, \mathcal{E}\cup\mathcal{E}^+)$, let $G_1=(V^{[1]}, E^{[1]})$ and $G_2=(V^{[2]}, E^{[2]})$ denote the two homogeneous networks within node sets $V^{[1]}=(v^{[1]}_1,\ldots,v^{[1]}_{n_1})$ and $V^{[2]}=(v^{[2]}_1,\ldots,v^{[2]}_{n_2}$), respectively. 
Furthermore, let $G_{12}=(V^{[1]}\cup V^{[2]}, \mathcal{E}^+)$ denote the bi-partite network formed between node sets $V^{[1]}$ and $V^{[2]}$. 
We subsequently refer to nodes in $V^{[1]}$ ($V^{[2]}$) as type-$[1]$ (type-$[2]$) nodes, edges in $E^{[1]}$ ($E^{[2]}$) as type-$[1]$ (type-$[2]$) edges, and edges in $E^{[12]}$ as type-$[12]$ edges.
We consider the following three matrices:
\begin{itemize}
\item $A^{[1]}$, the $n_1\times n_1$ 0-1 adjacency matrix of $G_1=(V^{[1]},E^{[1]})$, where $A^{[1]}_{ij}=1$ if and only if there is an edge between $v^{[1]}_i$ and $v^{[1]}_j$.
\item $A^{[2]}$, the $n_2\times n_2$ 0-1 adjacency matrix of $G_2=(V^{[2]},E^{[2]})$, where $A^{[2]}_{ij}=1$ if and only if there is an edge between $v^{[2]}_i$ and $v^{[2]}_j$.
\item $A^{[12]}$, the $n_1\times n_2$ 0-1 matrix of $G_{12}=(V^{[1]}\cup V^{[2]}, \mathcal{E}^+)$, where $A^{[12]}_{ij}=1$ if and only if there is an edge between $v^{[1]}_i$ and $v^{[2]}_j$.
\end{itemize}
Note that $A^{[12]}$ is not the adjacency matrix of $G_{12}=(V^{[1]}\cup V^{[2]}, \mathcal{E}^+)$, but only a submatrix of it. The adjacency matrix of $G_{12}$ is
\begin{equation*}
 \begin{pmatrix}
  \bm{0} & A^{[12]}  \\
  A^{[21]} & \bm{0} \\
 \end{pmatrix},
\end{equation*}
where $A^{[21]}=A{^{[12]}}^T$. We use $A^T$ to denote the transpose of matrix $A$. The matrix $A^{[12]}$ is usually referred to as the \textit{bi-adjacency matrix} of $G_{12}$. Since we only focus on networks with undirected edges, adjacency matrices $A^{[1]}$ and $A^{[2]}$ are both symmetric.
The heterogeneous network $\mathcal{G}$ can be uniquely represented by its $(n_1+n_2)\times(n_1+n_2)$ adjacency matrix $\mathcal{A}$,
\begin{equation*}
\mathcal{A}=
 \begin{pmatrix}
  A^{[1]} & A^{[12]}  \\
  A^{[21]} & A^{[2]} \\
 \end{pmatrix}.
\end{equation*}

\vspace{3pt}
\noindent {\bf 2.1. Null Model for Heterogeneous Networks}
\vspace{3pt}

The modularity function measures the difference between the observed network and the null model that characterizes networks with no community structure. 
To define the modularity function for a heterogeneous network, we need to formulate a null model for heterogeneous networks.

We introduce the following notations on degree sequences:
\begin{itemize}
\item $\mathbf{d}^{[1]}=(d^{[1]}_1,\ldots,d^{[1]}_{n_1})$, where $d^{[1]}_i=\sum_{j=1}^{n_1}{A}^{[1]}_{ij}$, $i=1,\ldots,n_1$, is the number of links incident to $v^{[1]}_i$ from $V^{[1]}$.
\item $\mathbf{d}^{[2]}=(d^{[2]}_1,\ldots,d^{[2]}_{n_2})$, where $d^{[2]}_i=\sum_{j=1}^{n_2}{A}^{[2]}_{ij}$, $i=1,\ldots,n_2$, is the number of links incident to $v^{[2]}_i$ from $V^{[2]}$.
\item $\mathbf{d}^{[12]}=(d^{[12]}_1,\ldots,d^{[12]}_{n_1})$, where $d^{[12]}_i=\sum_{j=1}^{n_2}{A}^{[12]}_{ij}$, $i=1,\ldots,n_1$, is the number of links incident to $v^{[1]}_i$ from $V^{[2]}$.
\item $\mathbf{d}^{[21]}=(d^{[21]}_1,\ldots,d^{[21]}_{n_2})$, where $d^{[21]}_i=\sum_{j=1}^{n_1}{A}^{[21]}_{ij}$, $i=1,\ldots,n_2$, is the number of links incident to $v^{[2]}_i$ from $V^{[1]}$.
\end{itemize}

From the definitions, we see that $\mathbf{d}^{[1]}$ is the vector of column (row) sums of $A^{[1]}$, $\mathbf{d}^{[12]}$ is the vector of row sums of $A^{[12]}$, $\mathbf{d}^{[21]}$ is the vector of column sums of $A^{[12]}$, and $\mathbf{d}^{[2]}$ is the vector of column (row) sums of $A^{[2]}$. 
Write the number of edges in $G_1$ as $m^{[1]}=\sum_{i=1}^{n_1}d^{[1]}_i$/2, the number of edges in $G_{12}$ as $m^{[12]}=\sum_{i=1}^{n_1}d^{[12]}_i$, and the number of edges in $G_2$ as $m^{[2]}=\sum_{i=1}^{n_2}d^{[2]}_i/2$. 
Define $\mathcal{D}=(\mathbf{d}^{[1]}, \mathbf{d}^{[12]}, \mathbf{d}^{[2]}, \mathbf{d}^{[21]})$.

An appropriate null model should satisfy the following two conditions. 
First, it should describe a random heterogeneous network with no community structure. 
Second, the networks from the null model should share basic structural properties with the observed network (Newman, 2006; Zhang and Chen, 2016). For the null model of a heterogeneous network, we propose to preserve the observed degree sequence $(\mathbf{d}^{[1]}, \mathbf{d}^{[12]}, \mathbf{d}^{[2]}, \mathbf{d}^{[21]})$. 
That is, the degrees $d_i^{[1]}$ and $d_i^{[12]}$ for each node $v^{[1]}_i$, $i=1\ldots,n_1$, are fixed. 
Similarly, the degrees $d_i^{[2]}$ and $d_i^{[21]}$ for each node $v^{[2]}_i$, $i=1\ldots,n_2$, are fixed. 

Preserving the observed degree sequence has been considered in various homogeneous network models in the literature (Chung and Lu, 2002; Newman and Girvan, 2004; Perry and Wolfe, 2012).
The edge distribution in real-world networks often displays high global inhomogeneity and local inhomogeneity. 
The global inhomogeneity refers to the feature that most nodes have low degrees while a few nodes have high degrees. The local inhomogeneity refers to the case that there is a high concentration of edges within certain groups of edges and low concentration of edges between these groups. The local inhomogeneity is also referred to as the community structure. 
To study the local inhomogeneity, it is important to control for the global inhomogeneity. That is, to study the community structure, it is important to control for the degree sequence.

The sample space in our null model is defined as
\begin{equation*}
\Sigma_{\mathcal{D}}=\{\mathcal{G}: \mathcal{G} \text{ is a simple heterogeneous network with degree sequence } \mathcal{D}\}. 
\end{equation*}
For a heterogeneous network $\mathcal{G}$ from the sample space, the null distribution is defined as 
\begin{equation}
p(\mathcal{G})=\frac{1}{|\Sigma_{\mathcal{D}}|}.
\label{null}
\end{equation}
Under the null model, every heterogeneous network from $\Sigma_{\mathcal{D}}$ is equally likely to occur and there is no preference for any network configuration. 
With the defined null model, we need to calculate the expectations $E_p(A_{ij}^{[1]})$, $E_p(A_{ij}^{[12]})$ and $E_p(A_{ij}^{[2]})$ for the modularity function defined in the Section 2.2. Here the expectation $E_p(\cdot)$ is taken with respect to $p(\cdot)$ in (\ref{null}).

To calculate $E(A_{ij}^{[l]})$ under the null model, we notice that
\begin{eqnarray*}
E(A_{ij}^{[l]})&=&\frac{|\Sigma_{\mathcal{D}|{A_{ij}^{[l]}=1}}|}{|\Sigma_{\mathcal{D}}|},
\end{eqnarray*}
where $\Sigma_{\mathcal{D}|{A_{ij}^{[l]}=1}}$ is the set of all simple heterogeneous networks in $\Sigma_{\mathcal{D}}$ with $A_{ij}^{[l]}$=1, $l=1,2$.
Denote $\Sigma_{\mathbf{d}^{[1]}}$ as the set of all simple homogeneous graphs with degree sequence $\mathbf{d}^{[1]}$, 
$\Sigma_{\mathbf{d}^{[2]}}$ as the set of all simple homogeneous graphs with degree sequence $\mathbf{d}^{[2]}$, 
and $\Sigma_{\mathbf{d}^{[12]},\mathbf{d}^{[21]}}$ as the set of all bipartite graphs with degree sequence $\mathbf{d}^{[12]}$ for type-$[1]$ nodes and degree sequence $\mathbf{d}^{[21]}$ for type-$[2]$ nodes.
We have $|\Sigma_{\mathcal{D}}|=|\Sigma_{\mathbf{d}^{[1]}}|\times|\Sigma_{\mathbf{d}^{[2]}}|\times|\Sigma_{\mathbf{d}^{[12]},\mathbf{d}^{[21]}}|$. 
It is easy to see that
\begin{eqnarray}
E(A_{ij}^{[l]})&=&\frac{|\Sigma_{\mathbf{d}^{[l]}|{A_{ij}^{[l]}=1}}|}{|\Sigma_{\mathbf{d}^{[l]}}|}, \quad l=1,2,
\label{e1}
\end{eqnarray}
where $|\Sigma_{\mathbf{d}^{[l]}|{A_{ij}^{[l]}=1}}|$ is the total number of simple homogeneous networks with degree sequence $\mathbf{d}^{[l]}$ and a link between nodes $i$ and $j$.
Similarly, we can show that
\begin{equation}
E(A_{ij}^{[12]})=\frac{|\Sigma_{\mathbf{d}^{[12]},\mathbf{d}^{[21]}|{A_{ij}^{[12]}=1}}|}{|\Sigma_{\mathbf{d}^{[12]},\mathbf{d}^{[21]}}|},
\label{e2}
\end{equation}
and
\begin{equation}
E(A_{ij}^{[21]})=\frac{|\Sigma_{\mathbf{d}^{[12]},\mathbf{d}^{[21]}|{A_{ij}^{[21]}=1}}|}{|\Sigma_{\mathbf{d}^{[12]},\mathbf{d}^{[21]}}|},
\label{e3}
\end{equation}
where $|\Sigma_{\mathbf{d}^{[12]},\mathbf{d}^{[21]}|{A_{ij}^{[12]}=1}}|$ is the total number of bi-partite graphs with degree sequences $\mathbf{d}^{[12]}$ for type-$[1]$ nodes, $\mathbf{d}^{[21]}$ for type-$[2]$ nodes and a link between the $i$th node of type-$[1]$ and the $j$th node of type-$[2]$.

Calculating the numerators and the denominators in (\ref{e1}), (\ref{e2}) and (\ref{e3}) is a difficult problem. 
Bender and Canfield (1978) and Bollob\'{a}s and McKay (1986) derived asymptotic formulas for the number of simple graphs with a fixed degree sequence and pre-specified structure zeroes (a structure zero at $A_{ij}$ means no edge can be placed between node $i$ and node $j$). 
Based on these asymptotic formulas, we have the following approximations for the expectations.
\textcolor{black}{
\begin{theorem}
Define $d_{max}^{[l]}=\max_{i=1}^{n_l}d_i^{[l]}$, $l=1, 2$, $d_{max}^{[12]}=\max_{i=1}^{n_1}d_i^{[12]}$, $d_{max}^{[21]}=\max_{i=1}^{n_2}d_i^{[21]}$, and assume that $n_1\le n_2$. Suppose for some $\eta>0$, $d_{max}^{[l]}\le (\log n_l)^{1/3}$, $m^{[l]}\ge\max\{\eta d_{max}^{[l]},(1+\eta)n_l\}$, $l=1, 2$, $d_{max}^{[12]}\le(\log n_1)^{1/3}$, $d_{max}^{[21]}\le(\log n_1)^{1/3}$, and $m^{[12]}\ge\max\{(2+\eta)n_2,\eta d_{max}^{[12]}, \eta d_{max}^{[21]}\}$.
Then $E(A_{ij}^{[l]})$ is 
$$
\frac{d^{[l]}_id^{[l]}_j}{2m^{[l]}}+o(e^{(\log n_l)^{4/5}}/n_l),\quad l=1,2,
$$
and $E(A_{ij}^{[12]})$ is  
$$
\frac{d_i^{[12]}d_j^{[21]}}{m^{[12]}}+o(n_2^{-3/4}).
$$
\end{theorem}
We refer to the online supplementary material for the proof. 
The conditions in Theorem 1 describe the density of the network as the network size tends to infinity. Specifically, the conditions $d_{max}^{[l]}\le (\log n_l)^{1/3}$, $l=1, 2$, $d_{max}^{[12]}\le(\log n_1)^{1/3}$ and $d_{max}^{[21]}\le(\log n_1)^{1/3}$ characterize the rates that the maximum node degrees increase at; these conditions are to make sure the network does not become extremely dense as the network size grows. The conditions $m^{[l]}\ge\max\{\eta d_{max}^{[l]},(1+\eta)n_l\}$, $l=1, 2$ and $m^{[12]}\ge\max\{(2+\eta)n_2,\eta d_{max}^{[12]}, \eta d_{max}^{[21]}\}$ provide lower bounds for edge sums $m^{[1]}$, $m^{[2]}$ and $m^{[12]}$(=$m^{[21]}$); these conditions are to make sure the network does not become extremely sparse as the network size grows.}

\textcolor{black}{The results in Theorem 1 indicate that $E(A_{ij}^{[l]})$ can be well approximated by $\frac{d^{[l]}_id^{[l]}_j}{2m^{[l]}}$ and $E(A_{ij}^{[12]})$ can be well approximated by $\frac{d_i^{[12]}d_j^{[21]}}{m^{[12]}}$. As such, we use these approximations in the modularity function defined in the next section.}

\vspace{3pt}
\noindent {\bf 2.2. Modularity Function}
\vspace{3pt}

We first consider heterogeneous networks with only two types of nodes ($L=2$). Later in this section, we generalize the results to heterogeneous networks with any $L\ge2$.
We define the $(n_1+n_2)\times(n_1+n_2)$ modularity matrix $\mathcal{M}$ for the heterogeneous network $\mathcal{G}$ as  
\begin{equation*}
\mathcal{M}=
 \begin{pmatrix}
  M^{[1]}/2m^{[1]} & M^{[12]}/m^{[12]}  \\
  M^{[21]}/m^{[21]} & M^{[2]}/2m^{[2]} \\
 \end{pmatrix},
\end{equation*}
where $M^{[1]}=A^{[1]}-E(A^{[1]})$, $M^{[2]}=A^{[2]}-E(A^{[2]})$, $M^{[12]}=A^{[12]}-E(A^{[12]})$ and $M^{[21]}=A^{[21]}-E(A^{[21]})$. 
If there are no edges within the type-$[1]$ (or type-$[2]$) nodes, we set $M^{[1]}/2m^{[1]}=\textbf{0}_{n_1\times n_1}$ (or $M^{[2]}/2m^{[2]}=\textbf{0}_{n_2\times n_2}$). 
Similarly, if there are no edges between type-$[1]$ and type-$[2]$ nodes, we set $M^{[12]}/m^{[12]}=\textbf{0}_{n_1\times n_2}$ and $M^{[21]}/m^{[21]}=\textbf{0}_{n_2\times n_1}$.

Define a 0-1 assignment matrix $\mathcal{B}$ of dimension $(n_1+n_2)\times K$ as
\begin{equation}
\mathcal{B}=
 \begin{pmatrix}
  B^{[1]}   \\
  B^{[2]} \\
 \end{pmatrix},
 \label{assign}
\end{equation}
where $B^{[1]}$ is an $n_1\times K$ matrix with $B^{[1]}_{ij}=1$ if node $v^{[1]}_i$ is in the $j$-th community and 0 otherwise, and $B^{[2]}$ is an $n_2\times K$ matrix with $B^{[2]}_{ij}=1$ if node $v^{[2]}_i$ is in the $j$-th community and 0 otherwise. 
The modularity function of a heterogeneous network is defined as
\begin{align}
Q(\mathcal{B},\mathcal{G})&=\frac{1}{4}tr(\mathcal{B}^T\mathcal{M}\mathcal{B})\\
&=\frac{1}{4}\left[\frac{1}{2m^{[1]}}tr(B^{{[1]}^T}M^{[1]}B^{[1]})+\frac{2}{m^{[12]}}\times tr(B^{{[1]}^T}M^{[12]}B^{[2]})+\frac{1}{2m^{[2]}}tr(B^{{[2]}^T}M^{[2]}B^{[2]})\right] \nonumber,
\end{align}
where $tr(\cdot)$ denotes the trace of a square matrix.
With some calculations, we can derive that 
$$
\frac{1}{2m^{[1]}}tr(B^{{[1]}^T}M^{[1]}B^{[1]})=\frac{1}{2m^{[1]}}\sum_{i,j}[A_{ij}^{[1]}-E(A_{ij}^{[1]})]I(B^{[1]}_{i.}=B^{[1]}_{j.}),
$$
$$
\frac{2}{m^{[12]}}tr(B^{{[1]}^T}M^{[12]}B^{[2]})=\frac{2}{m^{[12]}}\sum_{i,j}[A_{ij}^{[12]}-E(A_{ij}^{[12]})]I(B^{[1]}_{i.}=B^{[2]}_{j.}),
$$
$$
\frac{1}{2m^{[2]}}tr(B^{{[2]}^T}M^{[2]}B^{[2]})=\frac{1}{2m^{[2]}}\sum_{i,j}[A_{ij}^{[2]}-E(A_{ij}^{[2]})]I(B^{[2]}_{i.}=B^{[2]}_{j.}).
$$
Here $B_{i.}$ denotes the $i$th row of matrix $B$ and $I(\cdot)$ is an indicator function. 
For example, $I(B^{[1]}_{i.}=B^{[1]}_{j.})=1$ only when nodes $i$ and $j$ are both of type-$[1]$ and they are in the same community.
The first component $tr(B^{{[1]}^T}M^{[1]}B^{[1]})/2m^{[1]}$ and the third component $tr(B^{{[2]}^T}M^{[2]}B^{[2]})/2m^{[2]}$ calculate the differences between the observed number of intra-community edges and the expected number of intra-community edges in networks $G_1$ and $G_2$, respectively. The second component $tr(B^{{[1]}^T}M^{[12]}B^{[2]})/m^{[12]}$ calculates the difference between the observed number of intra-community edges and the expected number of intra-community edges in the bi-partite network $G_{12}$.

From the definition, we can see the modularity function $Q(\mathcal{B},\mathcal{G})\in[-1,1]$. 
When $Q(\mathcal{B},\mathcal{G})$ approaches 1, the observed numbers of type-$[1]$, type-$[2]$ and type-$[12]$ intra-community edges are much higher than the expected values, which indicates a strong community structure. 
On the other hand, when $Q(\mathcal{B},\mathcal{G})$ approaches 0, the observed numbers of type-$[1]$, type-$[2]$ and type-$[12]$ intra-community edges are close to the expected values, which indicates a weak community structure.

To generalize the modularity function to a heterogeneous network with $L$ types of nodes, we denote the adjacency matrix of $G_i(V^{[i]}, E^{[i]})$ as $A^{[i]}$ and the bi-adjacency matrix of $G_{ij}(V^{[i]}\cup V^{[j]}, E^{[ij]})$ as $A^{[ij]}$, $1\le i\neq j\le L$. 
Write the number of nodes in each type as $n_i=|V^{[i]}|$, $i=1,\ldots,L$. Further, write the number of edges in $G_i(V^{[i]}, E^{[i]})$ as $m^{[i]}$ and the number of edges in $G_{ij}(V^{[i]}\cup V^{[j]}, E^{[ij]})$ as $m^{[ij]}$, $1\le i\neq j\le L$.
The modularity function is defined as
\begin{equation}
Q(\mathcal{B},\mathcal{G})=\frac{1}{L^2}tr(\mathcal{B}^T\mathcal{M}\mathcal{B}),
\label{mod}
\end{equation}
where 
\begin{equation*}
\mathcal{M}=
\begin{pmatrix}
  M^{[1]}/2m^{[1]} & \ldots & M^{[1L]}/m^{[1L]}  \\
  \vdots   & \ddots & \vdots  \\
  M^{[L1]}/m^{[L1]} &\ldots & M^{[L]}/2m^{[L]} \\
\end{pmatrix},
\end{equation*}
and
\begin{equation*}
\mathcal{B}=
 \begin{pmatrix}
  B^{[1]}   \\  
  \vdots\\   
  B^{[L]} \\
 \end{pmatrix}.
\end{equation*}
Here $M^{[i]}=A^{[i]}-E(A^{[i]})$, $M^{[ij]}=A^{[ij]}-E(A^{[ij]})$, $1\le i\neq j\le L$. Matrix $\mathcal{B}$ is a $(n_1+\cdots+n_L)\times K$ assignment matrix defined similarly as that in (\ref{assign}). The expectations in the modularity function are approximated using the following corollary.
\textcolor{black}{
\begin{cor}
Define $d_{max}^{[l]}=\max_{i=1}^{n_l}d_i^{[l]}$, $l=1, \ldots, L$, $d_{max}^{[l_1l_2]}=\max_{i=1}^{n_{l_1}}d_i^{[l_1l_2]}$, $1\le l_1\neq l_2\le L$, and assume that $n_1\le\cdots\le n_L$.
Suppose for some $\eta>0$, $d_{max}^{[l]}\le (\log n_l)^{1/3}$, $m^{[l]}\ge\max\{\eta d_{max}^{[l]},(1+\eta)n_l\}$, $l=1, \ldots, L$, $d_{max}^{[l_1l_2]}\le(\log n_{1})^{1/3}$, and $m^{[l_1l_2]}\ge\max\{(2+\eta)n_L,\eta d_{max}^{[l_1l_2]}, \eta d_{max}^{[l_2l_1]}\}$.
Then $E(A_{ij}^{[l]})$ is 
$$
\frac{d^{[l]}_id^{[l]}_j}{2m^{[l]}}+o(e^{(\log n_l)^{4/5}}/n_l),\quad l=1,\ldots,L,
$$
and $E(A_{ij}^{[l_1l_2]})$ is 
$$
\frac{d_i^{[l_1l_2]}d_j^{[l_2l_1]}}{m^{[l_1l_2]}}+o(n_L^{-3/4}),\quad 1\le l_1\neq l_2\le L.
$$
\end{cor}
}
The corollary is directly available from Theorem 1. Since a larger modularity value indicates a stronger community structure, the community assignment of nodes in the heterogeneous network $\mathcal{G}$ is identified by maximizing the modularity function with respect to $\mathcal{B}$.  In the next section, we introduce a Louvain type method for efficient modularity function maximization.

\vspace{3pt}
\setcounter{chapter}{3}
\setcounter{equation}{0}
\noindent {\bf 3. Modularity Maximization}
\vspace{3pt}

Our goal is to find the community assignment matrix $\mathcal{B}$ that maximizes the modularity function in (\ref{mod}), i.e.,
$$
\argmax_{\substack{{\mathcal{B}_{(n_1+\cdots+n_L)\times K}}\\{K\in \mathbb{Z}^+}}} tr(\mathcal{B}^T\mathcal{M}\mathcal{B}).
$$
Maximizing this objective function is a very difficult problem, especially since the number of communities $K$ is generally unknown. Brandes et al. (2008) showed that finding the partition that maximizes the modularity function for a homogeneous network is NP-hard. Existing heuristic approaches for maximizing the modularity function come from various fields, including computer science, physics and sociology (Clauset et al., 2004; Massen and Doye, 2005; Newman, 2006; Reichardt and Bornholdt, 2006; Agrawal and Kempe, 2008). 
In this paper, we adopt a Louvain type maximization method.

The Louvain maximization method is a hierarchical clustering method proposed in Blondel et al. (2008). 
The technique was developed to maximize the modularity function of a homogeneous network. 
The optimization procedure is carried out in two phases that are repeated iteratively. 
The first phase starts by assigning each node in the network to its own community (each community contains one and only one node). 
Then each node $i$ is moved to the neighboring community that results in the largest increase in modularity (if no increase is possible, then node $i$ remains in its original community). A neighboring community of node $i$ is defined as a community that node $i$ is linked to.
In the second phase, the algorithm aggregates nodes in the same community and ``constructs" a new network whose nodes are the communities from the first phase. 
The edges between the new nodes are calculated using the edges connecting the two corresponding communities (see Blondel et al. (2008) for details). 
These steps are repeated iteratively until the modularity reaches its local maximum.

The Louvain method were successfully applied to various homogeneous networks of sizes up to 100 million nodes and billions of links. Using the Louvain method for community detection in a typical network with 2 million nodes only takes several minutes on a standard PC (Blondel, 2011). Fortunato (2012) noted that the modularity maximum found by the Louvain method often compares favorably with those found by using the methods in Clauset et al. (2004) and Wakita and Tsurumi (2007).

Similar to the Louvain method, finding the maximizer of the proposed heterogeneous network modularity function can also be carried out in two phases that are repeated iteratively. 
To ease the presentation, we focus on the case where $L=2$, i.e., there are two types of nodes. First we define a term ``unit". 
A unit may contain one node of any type or two nodes of different types. A community consists of several units.
To initialize, we assign each node in the network to its own unit. 
Therefore, if there are $n_1$ type-$[1]$ nodes and $n_2$ type-$[2]$ nodes, the algorithm starts with $n_1+n_2$ units. 
In the first phase, we start by assigning each unit to its own community. Then we calculate the change in modularity when unit $i$ is assigned to each one of its neighboring communities. 
A neighboring community of unit $i$ is defined as a community that unit $i$ is linked to.
Once this value is calculated for every community that unit $i$ is linked to, we assign unit $i$ to the community that leads to the largest increase in modularity. 
If no move increases the modularity, unit $i$ remains in its original community. 
This step is applied repeatedly to the units in the network until no increase in modularity can be achieved.
In the second phase, we examine each community from the first phase and merge nodes of the same type in each community. 
This community then becomes a new unit in the next step. 
If two communities are linked, then there is an edge between the two new units; if two communities are not linked, then there is no edge between the two new units.
We repeat these two phases iteratively until there is no move possible and the modularity reaches a local maximum. 

\begin{figure}[!htb]
 \centering
\includegraphics[scale=0.55,trim=5mm 0 0 0]{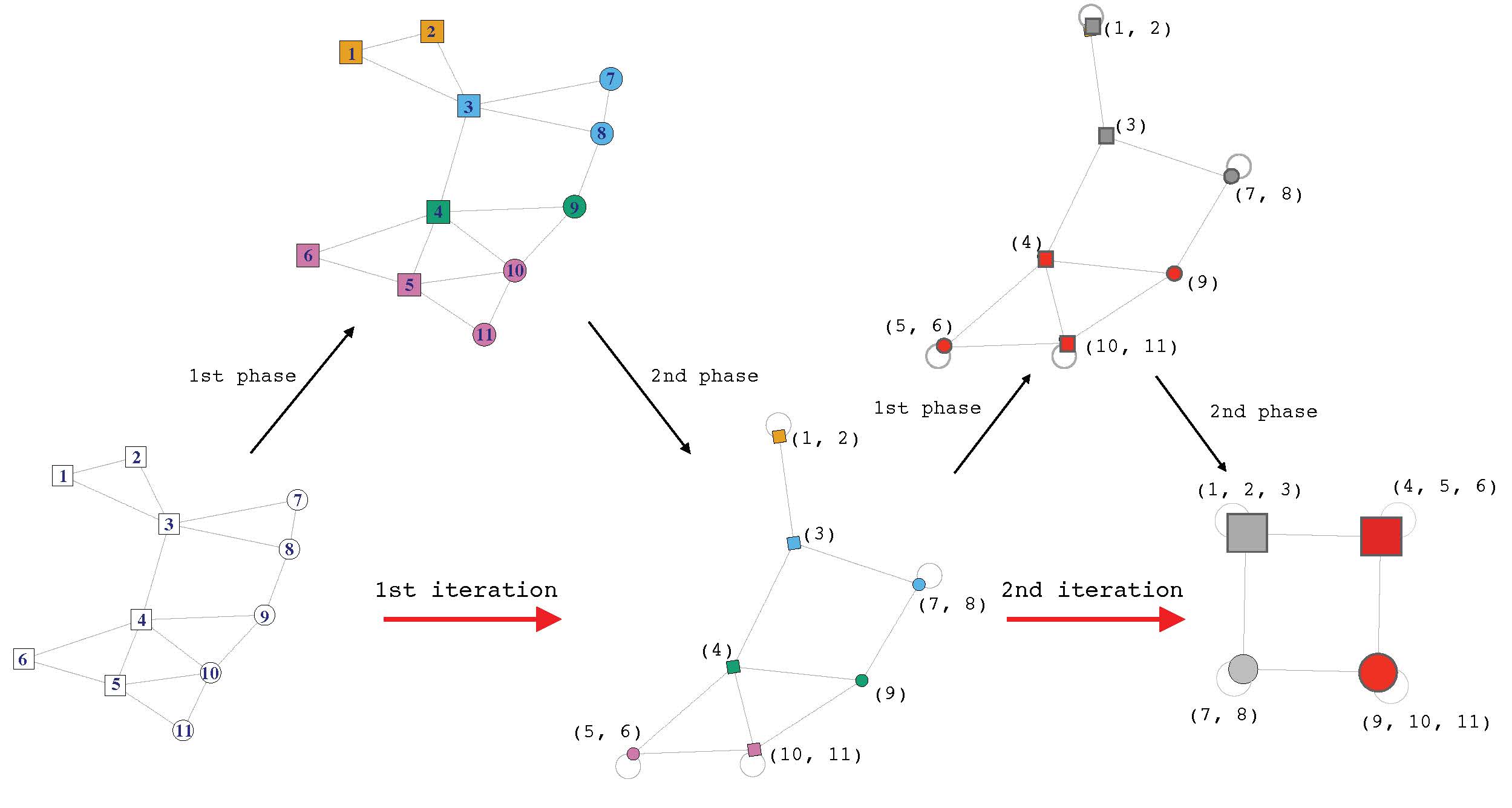}
\caption{\small A visualization of the steps in the proposed algorithm. The two types of nodes are represented by squares and circles, respectively. Nodes of the same color are in the same community. After the first iteration and the second iteration, each node in the graph has a bracket next to it indicating the nodes from the original graph it contains. }
\label{flow}
\end{figure}

As an example, Figure~\ref{flow} shows the application of the proposed algorithm to a heterogeneous network with 2 types of nodes. Each iteration contains two phases. 
In the first iteration, the number of communities changes from 11 to 4. 
After the first iteration, nodes 1 and 2 are merged and treated as one node, say $v^*_{1,2}$, in the second iteration; similarly, nodes 7 and 8 are merged and treated as one node, say $v^*_{7,8}$; node 3 does not merge with any node and is treated as one node, say $v^*_3$. In the second iteration, nodes $\{v^*_3, v^*_{7,8}\}$ form a unit and node $v_{1,2}^*$ is a unit. 
During the first phase in the second iteration, we compute the change in modularity when we place unit $v^*_{1,2}$ and unit $\{v^*_3, v^*_{7,8}\}$ in one community. 
If the modularity increases, we place $v^*_{1,2}$ and $\{v^*_3, v^*_{7,8}\}$ in one community; if the modularity decreases, the two units remain in their original communities. 
In the second iteration, the number of communities changes from 4 to 2. The algorithm outputs two communities with the first community including nodes 1, 2, 3, 7, 8 and the second community including nodes 4, 5, 6, 9, 10, 11.

The algorithm can be summarized as follows.
\begin{algorithm}
\text{}
Take the modularity matrix $\mathcal{M}$ as input:
\begin{description}
\item{1.} Assign each node to its own unit.
\item{2.} Assign each unit to its own community.
\item{3.} For each unit $i$, place it into the neighboring community that leads to the largest modularity increase. If no such move is possible, unit $i$ remains in its current community.
\item{4.} Apply Step 3 repeatedly to the units in the network until no units can be moved.
\item{5.} If the modularity is higher than the modularity from the previous iteration, then merge the nodes of the same type in each community; each community is treated as a unit and return to Step 2. If not, output the community assignment and the modularity value from the previous iteration. 
\end{description}
\end{algorithm}

The result of the algorithm depends on the initial ordering of the nodes.
In addition, in Step 3, each node is assigned to the community that leads to the largest modularity increase. 
If there are several communities that all lead to the largest increase, one community is randomly selected. 
Hence, the Louvain method may not arrive at the same result in successive runs.
In our analysis, we apply the Louvain method $\kappa$ times with random node orderings to find the maximum of the modularity function.
In general, $\kappa$ should increase with the size and the complexity of the network. 
In our simulation and real data analysis, we set $\kappa=100$. We do not observe notable improvements in the maximized modularity function for $\kappa>100$. However, other networks of comparable or larger sizes may benefit if larger values of $\kappa$ are selected.

In the implementation of the Louvain method, the decision of whether and where to move a node can be computed in $O(1)$ time. Consequently, the complexity per iteration is $O(m)$, where $m$ is the total number of edges in the network. 
An upper bound on the total running time of the algorithm can be calculated as $O(rm)$, where $r$ is the total number of iterations. 
A trivial upper bound on $r$, which gives the worst case, is $O(m^2)$.
While no nontrivial upper bound has been established on the number of iterations, the method needs only tens of iterations in practice to converge.

We note that the Louvain maximization method does not require the number of communities to be pre-specified. 
In cases where it is desirable to fix the number of communities at $K^*$ in the procedure, the Louvain method can still be applied. 
Specifically, if $K^*$ is reached during the iterations in the algorithm, we would stop the procedure and output the community assignment. 
If $K^*$ is not reached after the algorithm finishes, i.e., the algorithm finds $K>K^*$, then we would continue with the algorithm and stop once $K^*$ is reached; when continuing the algorithm, a small modification is that in Step 3, we would move unit $i$ into the neighboring community that leads to the smallest modularity decrease.

\vspace{3pt}
\setcounter{chapter}{4}
\setcounter{equation}{0}
\noindent {\bf 4. Consistency}
\vspace{3pt}

The consistency of community detection approaches for homogeneous networks has been studied extensively (Bickel and Chen, 2009; Rohe et al., 2011; Zhao et al., 2012; Jin, 2015). 
However, theoretical properties of community detection methods for heterogeneous networks are largely unaddressed. 
In this section, we investigate the consistency property of our proposed method under a heterogeneous stochastic blockmodel framework.
The consistency property of our method when applied to bipartite networks or multipartite networks follow as special cases.

Consider a heterogeneous network $\mathcal{G}=(\bigcup_{i=1}^LV^{[i]},\mathcal{E}\cup\mathcal{E}^+)$ with latent community labels $\bc^{[l]}=(c_1^{[l]}, \ldots,c_{n_l}^{[l]})$, $l=1,\dots,L$, where $c_i^{[l]}\in\{1,\ldots,K\}$ is the community that the $i$th node of type-$[l]$ belongs to. 
Write $\mathcal{C}=(\bc^{[1]},\ldots,\bc^{[L]})$ and $n=\sum_{l=1}^Ln_l$. 
We assume that the sizes of $V^{[l]}$, $l=1,\ldots,L$, are balanced, i.e., $\min_{l}n_l/n$ is bounded away from zero. 
We define a community detection criterion $F(\mathcal{C},\mathcal{G})$ to be consistent if 
$$
\mathcal{\hat C}=\arg\max_\mathcal{C} F(\mathcal{C},\mathcal{G})
$$
satisfies
$$
\forall\epsilon>0,\quad P\left[\frac{1}{n}\sum_{l=1}^L\sum_{i=1}^{n_l}I(\hat c_i^{[l]}\neq c_i^{[l]})<\epsilon\right]\rightarrow1 \text{ as}\quad n\rightarrow\infty.
$$
This definition of consistency is a generalization of the one proposed in Zhao et al. (2012) for homogeneous networks. The definition requires that the error rate tends to zero in probability as the number of nodes goes to infinity.

Next we introduce the heterogeneous stochastic blockmodel, which serves as the framework of our theoretical development. 
Consider a heterogeneous network $\mathcal{G}=(\bigcup_{i=1}^LV^{[i]},\mathcal{E}\cup\mathcal{E}^+)$ with latent community label $\mathcal{C}$. 
Write the adjacency matrix of $G_l(V^{[l]},E^{[l]})$ as $A^{[l]}$, $l=1,\dots,L$, and the bi-adjacency matrix of $G_{l_1l_2}(V^{[l_1]}\bigcup V^{[l_2]},E^{[l_1l_2]})$ as $A_{ij}^{[l_1l_2]}$, $1\le l_1\neq l_2\le L$.
In a heterogeneous stochastic blockmodel, each $A_{ij}^{[l]}$ is an independent Bernoulli random variable with
$$
E(A_{ij}^{[l]}\mid c^{[l]}_i=a,c^{[l]}_j=b)=P_{ab}^{[l]},
$$
and each $A_{ij}^{[l_1l_2]}$ is an independent Bernoulli random variable with
$$
E(A_{ij}^{[l_1l_2]}\mid c^{[l_1]}_i=a,c^{[l_2]}_j=b)=P_{ab}^{[l_1l_2]},
$$
where $P^{[l]}$ is a symmetric $K\times K$ probability matrix specifying the connecting probabilities between different communities of type-$[l]$ nodes, and $P^{[l_1l_2]}$ is a $K\times K$ probability matrix specifying the connecting probabilities between type-$[l_1]$ nodes and type-$[l_2]$ nodes in different communities. Note that by definition, we have $P^{[l_1l_2]}={P^{[l_2l_1]}}'$. Define $\pi^{[l]}=(\pi^{[l]}_1,\ldots,\pi^{[l]}_K)$ where $\pi^{[l]}_k=\frac{1}{n}\sum_{i=1}^{n_l}I(c^{[l]}_i=k)$, $l=1,\ldots,L$.

To ensure sparsity, the entries in the probability matrices need to tend to zero as the network grows in size. Otherwise, the network is going to become unrealistically dense. Following Bickel and Chen (2009), we define the expected degree $\lambda_n=n\rho_n$, where $\rho_n\equiv P(Edge)\rightarrow0$. We can reparameterize $P^{[l]}$ as $\tilde P^{[l]}=\rho_nP^{[l]}$, where $P^{[l]}$ is fixed as $n\rightarrow\infty$. This reparameterization allows us to separate $\rho_n$ from the structure of the network. See Bickel and Chen (2009) for a more detailed discussion of the reparameterization.

Consider the modularity function $Q(\mathcal{B},\mathcal{G})$ in (\ref{mod}). The assignment matrix $\mathcal{B}$ and the assignment vector $\mathcal{E}=(\be^{[1]},\ldots,\be^{[L]})$ with $\be^{[l]}=(e_1^{[l]}, \ldots,e_{n_l}^{[l]})$, $l=1,\dots,L$, have one to one correspondence. To simplify the notations, we write the modularity function $Q(\mathcal{B},\mathcal{G})$ as $Q'(\mathcal{E},\mathcal{G})$ in this section. The consistency property of the proposed heterogeneous network community detection criterion $Q'(\mathcal{E},\mathcal{G})$ is introduced in the following theorem.

\begin{theorem}
\label{th2}
Consider $\mathcal{G}(\bigcup_{i=1}^LV^{[i]},\mathcal{E}\cup\mathcal{E}^+)$ from a heterogeneous stochastic blockmodel with parameters $P^{[l]}$ and $P^{[l_1l_2]}$, $l=1,\dots,L$, $1\le l_1\neq l_2\le L$. Define 
$$
T_{ab}^{[l]}=\frac{\pi^{[l]}_a\pi^{[l]}_bP^{[l]}_{ab}}{\sum_{ab}\pi^{[l]}_a\pi^{[l]}_bP^{[l]}_{ab}},\quad\text{and}\quad T_{ab}^{[l_1l_2]}=\frac{\pi^{[l_1]}_a\pi^{[l_2]}_bP^{[l_1l_2]}_{ab}}{\sum_{ab}\pi^{[l_1]}_a\pi^{[l_2]}_bP^{[l_1l_2]}_{ab}}.
$$
Write $W^{[l]}=T^{[l]}-(T^{[l]}\bm{1})(T^{[l]}\bm{1})'$ and $W^{[l_1l_2]}=T^{[l_1l_2]}-(T^{[l_1l_2]}\bm{1})(T^{[l_1l_2]}\bm{1})'$.
If the parameters satisfy
\begin{equation}
\label{condition}
\sum_{l=1}^LW^{[l]}_{aa}+\sum_{l_1\neq l_2}^LW^{[l_1l_2]}_{aa}>0,\quad \sum_{l=1}^LW^{[l]}_{ab}+\sum_{l_1\neq l_2}^LW^{[l_1l_2]}_{ab}<0 \quad\text{for all}\quad a\neq b,
\end{equation}
then the proposed modularity function $Q'(\mathcal{E},\mathcal{G})$ is consistent as $\lambda_n\rightarrow\infty$. 
\end{theorem}

We refer to the online supplementary material for the proof. 
This result on consistency suggests that if networks are from a heterogeneous stochastic blockmodel with $K$ communities, the community labels obtained from maximizing the modularity function $Q'(\mathcal{E},\mathcal{G})$ will approach the true community labels as the number of nodes goes to infinity.

Conditions (\ref{condition}) in Theorem~\ref{th2} essentially require that, on average,
edges are more likely to be established within communities than between communities, even though
community structures may not exist in all different types of edges. 
One example is the parameters in Simulation Setting 3 (see Section 5). In that case, edges within type-$[1]$ or type-$[2]$ nodes have no community structure, but edges linking type-$[1]$ nodes and type-$[2]$ nodes have community structure .

In a homogeneous network ($L=1$) with $K=2$, the conditions in (\ref{condition}) can be simplified to 
$$
P^{[1]}_{11}P^{[1]}_{22}>(P^{[1]}_{12})^2,
$$
which is satisfied if $P^{[1]}_{11}>P^{[1]}_{12}$ and $P^{[1]}_{22}>P^{[1]}_{12}$; these conditions describe settings in which edges are more likely to be
established within communities than between communities.
In the case when $L=2$ and $K=2$, the conditions in (\ref{condition}) are satisfied if 
$$
P^{[1]}_{11}+P^{[2]}_{11}+P^{[12]}_{11}+P^{[21]}_{11}>P^{[1]}_{12}+P^{[2]}_{12}+P^{[12]}_{12}+P^{[21]}_{12}
$$
and 
$$
P^{[1]}_{22}+P^{[2]}_{22}+P^{[12]}_{22}+P^{[21]}_{22}>P^{[1]}_{12}+P^{[2]}_{12}+P^{[12]}_{12}+P^{[21]}_{12}.
$$
These conditions describe that, on average, edges are more likely to be established within communities.

\vspace{3pt}
\setcounter{chapter}{5}
\setcounter{equation}{0}
\noindent {\bf 5. Simulation Study}
\label{sec::sim}
\vspace{3pt}

In this section, we evaluate the performance of the proposed method through simulated heterogeneous networks, and compare it to the performances of the following methods:
\begin{itemize}
\item \textbf{Method 1}: treat the whole heterogeneous network as one homogeneous network, i.e., do not distinguish the different types of nodes and edges;
\item \textbf{Method 2}: decompose the heterogeneous network with $L$ different types of nodes into $L$ homogeneous networks and consider each homogeneous network separately, i.e., discard information from the edges linking different types of nodes. 
\end{itemize}
The community assignments from Method 1 and Method 2 are obtained through maximizing the modularity functions defined on the homogeneous networks (Newman and Girvan, 2004). 
\textcolor{black}{When finding communities using our proposed method, Method 1 and Method 2, we do not fix the number of the communities and treat it as an unknown quantity.}
 
The model used to generate heterogeneous networks has two types of nodes ($L=2$) and three communities ($K=3$). We consider a stochastic block model type of structure with the probability matrix given as
$$
P= 
 \begin{pmatrix}
  P^{[1]} & P^{[12]} \\
  P^{[21]} & P^{[2]} \\
\end{pmatrix},
 $$
 where
 \begin{eqnarray*}
P^{[1]}&=&p_1\textbf{1}_K\textbf{1}_K'+r_1\bm{I}_K,\\
P^{[2]}&=&p_2\textbf{1}_K\textbf{1}_K'+r_2\bm{I}_K,\\
P^{[12]}=P^{[21]}&=&p_3\textbf{1}_K\textbf{1}_K'+r_3\bm{I}_K,
\end{eqnarray*}
where $\textbf{1}_K$ is the $K$-vector of 1's and $\bm{I}_K$ is the $K$-by-$K$ identity matrix. 
Here $P^{[1]}$ is a $3\times 3$ probability matrix characterizing the connecting probabilities between type-$[1]$ nodes in the three communities. 
For example, $P^{[1]}_{22}$ is the probability that there is an edge between two type-$[1]$ nodes that are both in the second community. Similarly, $P^{[2]}$ is the probability matrix characterizing the connecting probabilities between type-$[2]$ nodes in the three communities; $P^{[12]}$ and $P^{[21]}$ are the probability matrices characterizing the connecting probabilities between nodes of different types in the three communities. In the type-$[1]$ (type-$[2]$) homogeneous network, $p_1$ ($p_2$) represents the inter-community connecting probability and $p_1+r_1$ ($p_2+r_2$) represents the intra-community connecting probability. 
In the type-$[1]$ to type-$[2]$ bipartite network, $p_3$ describes the inter-community connecting probability and $p_3+r_3$ describes the intra-community connecting probability. 
The strength of the community structure is therefore regulated by $r_1$, $r_2$ and $r_3$.

Our main goal in this simulation study is to study how the clustering results from the proposed method, Method 1 and Method 2 change with $r_3$ under different settings. A higher value for $r_3$ results in a stronger intra-community connection between a type-$[1]$ node and a type-$[2]$ node, i.e., more information is contained in the edges linking different types of nodes.

We consider three simulation settings in this section. In all three settings, we gradually increase $r_3$ and compare the performances of the proposed method, Method 1 and Method 2.
In Simulation 1, the two homogeneous networks of type-$[1]$ nodes and type-$[2]$ nodes both have weak community structures.
In Simulation 2, the homogeneous network of type-$[1]$ nodes has weak community structure and the homogeneous network of type-$[2]$ nodes has no community structure.
In Simulation 3, neither of the two homogeneous networks has a community structure.
We set the number of type-$[1]$ nodes to 600 and assign 200 nodes to each community; we set the number of type-$[2]$ nodes to $300$ and assign 100 nodes to each community.

Before we discuss the results from the simulations, we first introduce a numerical measure to quantify the difference between two partitions. In this work, we adopt the normalized mutual information measure (Danon et al., 2005). Consider the community assignment $\{x_i\}$ and $\{y_i\}$, where $x_i$ and $y_i$ indicate the cluster labels of vertex $i$ in partitions $\mathcal{X}$ and $\mathcal{Y}$, respectively. Assume that labels $x$ and $y$ are the observed values of two random variables $X$ and $Y$. The normalized mutual information (NMI) is measured by
\begin{equation*}
NMI(\mathcal{X},\mathcal{Y})=\frac{2I(X,Y)}{H(X)+H(Y)},
\end{equation*}
where $I(X,Y)=H(X)-H(X|Y)$ is the mutual information and $H(X)=-\sum_xP(x)\log P(x)$ is the Shannon entropy of $X$. The normalized mutual information equals 1 if the two partitions are identical, and its expected value is 0 if the two partitions are independent.

\begin{figure}[!htb]
\centering
\includegraphics[scale=0.6, trim=0 5mm 0 0]{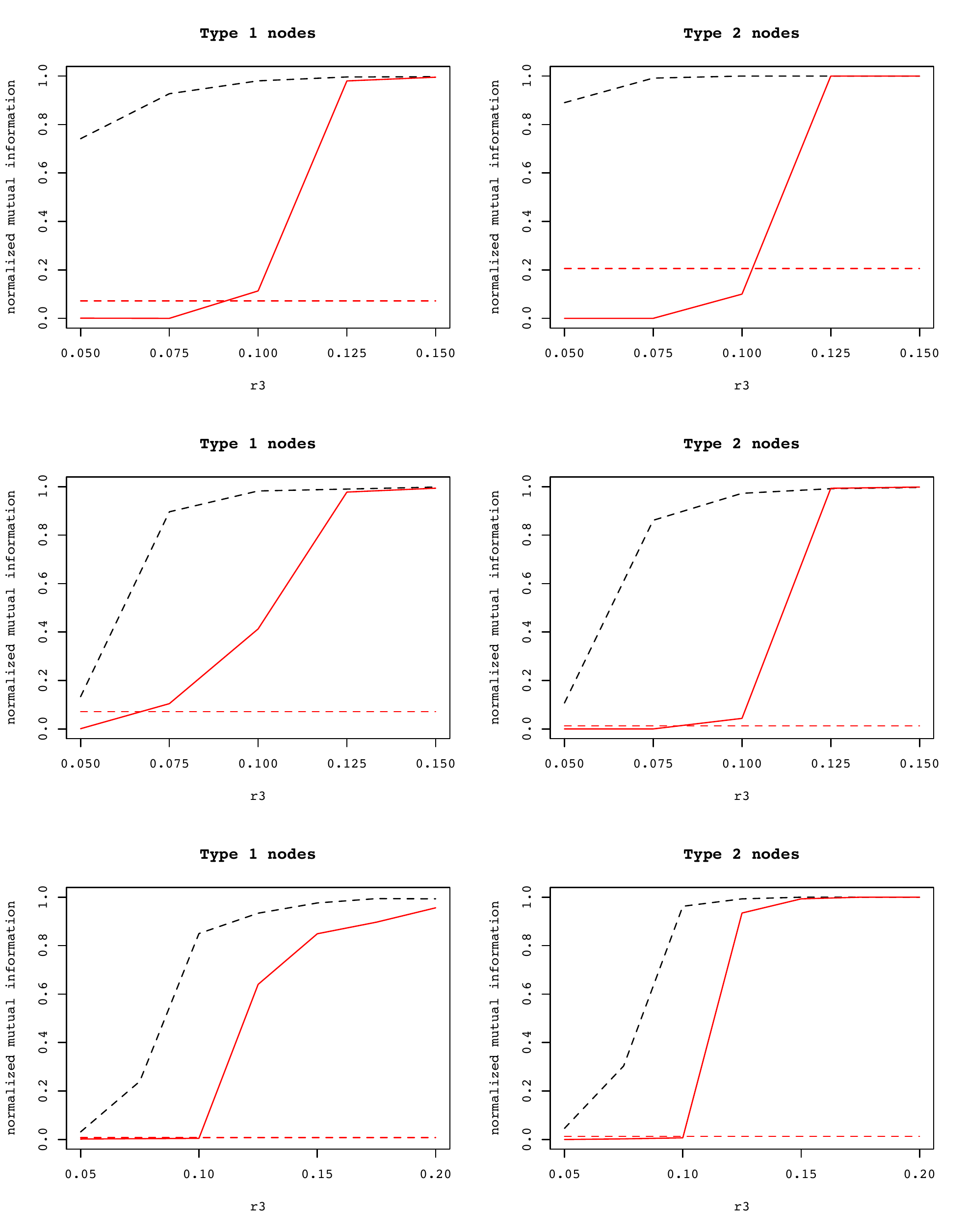}
\caption{\small Average NMI between the true community membership and the community membership obtained from the proposed method (black dashed line), Method 1 (red solid line) and Method 2 (red dashed line). Top panel: results from Simulation 1; middle panel: results from Simulation 2; bottom panel: results from Simulation 3.}
\label{sim}
\end{figure}

\noindent\textbf{Simulation Setting 1:}\\
In this simulation, we set the parameters $p_1=0.1$, $r_1=0.05$, $p_2=0.2$, $r_2=0.1$ and $p_3=0.05$. 
Under this setting, there are weak community structures within both type-$[1]$ nodes and type-$[2]$ nodes. 
We can see type-$[1]$ nodes and type-$[2]$ nodes behave quite differently; compared to type-$[1]$ nodes, type-$[2]$ nodes are much more densely connected amongst themselves. 
In this simulation, we gradually change $r_3$ from 0.05 to 0.15. 
For each $r_3$ value, we simulate 100 heterogeneous networks from the model. 
For each heterogeneous network, we apply the proposed method, Method 1, Method 2 and calculate the NMI between the obtained community detection results and the true community membership. 
The average of the NMI from the 100 simulations is summarized in the top panel of Figure~\ref{sim}.

We can see that the proposed method performs better than Method 1 and Method 2 on all values of $r_3$. 
Method 2 does not have satisfactory performance, with average NMI below 0.25 for both types of nodes. 
This is because the two homogeneous networks of type-$[1]$ nodes and type-$[2]$ nodes both have very weak community structures, and Method 2 does not take into consideration the edges linking different types of nodes. 
Note that our proposed method has good performance even when the connections between type-$[1]$ nodes and type-$[2]$ nodes display a weak community structure with $r_3=0.05$. 

\noindent\textbf{Simulation Setting 2:}\\
In this simulation, we set the parameters $p_1=0.1$, $r_1=0.05$, $p_2=0.2$, $r_2=0$, $p_3=0.05$.
Under this setting, the homogeneous network of type-$[2]$ nodes has no community structure. 
Similar to Simulation 1, we gradually increase $r_3$ from 0.05 to 0.15 and simulate 100 heterogeneous networks from the model. 
The average of the NMI from the 100 simulations is summarized in the middle panel of Figure~\ref{sim}.

We can see the proposed method outperforms Method 1 and Method 2 on all values of $r_3$. For type-$[2]$ nodes, the NMI from Method 2 is 0 since $r_2=0$. When $r_3=0.05$, the proposed method yields unsatisfactory performance. This is because the community structure is very weak within type-$[1]$ nodes, and between type-$[1]$ and type-$[2]$ nodes. When $r_3$ increases slightly to 0.075, we see notable improvement in the performance of the proposed method.

\noindent\textbf{Simulation Setting 3:}\\
In this simulation, we set the parameters $p_1=0.1$, $r_1=0$, $p_2=0.2$, $r_2=0$, $p_3=0.05$.
Under this setting, there are no community structures within type-$[1]$ nodes or type-$[2]$ nodes. 
In this simulation, we gradually increase $r_3$ from 0.05 to 0.20. The average of the NMI from 100 simulations between the true membership and the community membership calculated from the proposed method, Method 1 and Method 2 are summarized in the bottom panel of Figure~\ref{sim}.

We can see the proposed method has the best performance out of the three methods consistently. 
The NMI from Method 2 is 0 for both type-$[1]$ nodes and type-$[2]$ nodes, since there are no community structures within type-$[1]$ nodes or within type-$[2]$ nodes. 
For $r_3=0.05$ and 0.075, the proposed method yields NMI below 0.4 for both types of nodes. The low NMI is a result of the weak community structure in the simulated heterogeneous networks, with both $r_1$ and $r_2$ equal 0. When $r_3$ increases to 0.1, we see a significant improvement in the performance of the proposed method, while Method 1 still performs poorly.

\vspace{3pt}
\setcounter{chapter}{6}
\setcounter{equation}{0}
\noindent {\bf 6. Real Data Application}
\vspace{3pt}

\vspace{3pt}
\noindent {\bf 6.1. DBLP Dataset}
\vspace{3pt}

DBLP (Digital Bibliography \& Library Project) is a computer science bibliography website, listing more than 3.4 million journal articles, conference papers, and other publications in computer science.
 Gao et al. (2009) and Ji et al. (2010) extracted a connected subset of the DBLP dataset, containing bibliographical records from four research areas: database, data mining, information retrieval, and artificial intelligence. 
This network contains three types of nodes: paper, conference, and author. 
Among the three types of nodes, there are two types of edges: paper-conference (paper published at conference), paper-author (paper written by author).
This dataset consists of 14,376 papers written by 14,475 authors, and published at 20 conferences. 
Each one of the 20 conferences is labeled with the research area it covers. 
Each research area has five conferences.
The true research area is available for 4,057 authors that are connected to a subset of 14,328 papers, covering all 20 conferences. 
The objective in this real data application is to correctly identify the research areas of the authors. 
Since error rates can be calculated only for labeled authors, we focus our data analysis on this labeled subset of the data.

\textcolor{black}{Applying the proposed maximization method to the heterogeneous network modularity function with $K=4$ and $\kappa=100$, we cluster the heterogeneous network into four communities with the maximized modularity value 0.65. 
One application of the proposed maximization procedure takes less than 20 seconds on an iMac with 3.2 GHz Intel Core i5.
We label the research area of each community using the conferences each community contains (see Table~\ref{match}). 
The misclassification rate for the conferences is 0\%. 
We label the authors in each community with the research area that the community belongs to and compare the labels to the ground truth. 
The misclassification rate for the authors is 8.84\%. 
}

\begin{table}[!htb]
\centering
\begin{tabular}{c|c|l}
Community & Conferences                    & Research Area           \\ \hline
1         & PODS, ICDE, SIGMOD, EDBT, VLDB & Database                \\
2         & ICDM, PAKDD, PKDD, KDD, SDM    & Data Mining             \\
3         & AAAI, IJCAI, ECML, ICML, CVPR  & Artificial Intelligence \\
4         & WWW, WSDM, CIKM, ECIR, SIGIR   & Information Retrieval   \\ \hline
\end{tabular}
\caption{The conferences in each community and the research areas the conferences cover.}
\label{match}
\end{table}

\textcolor{black}{
We also considered Method 1 and Method 2 described in Section 5. 
Method 2 cannot be applied since there are no author-author, paper-paper or conference-conference connections.
Method 1 can be applied using the standard Louvain maximization approach with $K=4$. 
However, the identified communities are very difficult to interpret. 
For example, one community contains only papers and one community contains only conferences. 
This is not surprising since Method 1 treats author nodes, paper nodes and conference nodes equally, even though they behave differently in the DBLP network. 
}

\vspace{3pt}
\noindent {\bf 6.2. DBLP MovieLens Dataset}
\vspace{3pt}

MovieLens (https://movielens.org/) is a website that allows users to review movies; based on their reviews, users can receive personalized movie recommendations. 
The website was created in 1997 by a research lab in the Department of Computer Science and Engineering at the University of Minnesota to collect research data (Harper and Konstan, 2015). 
The MovieLens dataset (https://grouplens.org/datasets/movielens/) contains reviews from 943 users on 1,682 movies from 18 movie genre including action, adventure, animation, children's, comedy, crime, documentary, drama, fantasy, film-noir, horror, musical, mystery, romance, sci-fi, thriller, war and western. 
Using the MovieLens dataset, we construct a heterogeneous network with three types of node: user, movie and genre, and two types of edges: user-movie (movie reviewed by user), movie-genre (movie contained in genre). 
The objective in this real data application is to find communities in this heterogeneous network. The identified communities can be used to classify movies, users, and make movie recommendations since users in the same community are more likely to watch the movies contained in this community.

Applying the proposed heterogeneous network community detection technique with $\kappa=100$, we identified 7 communities with maximized modularity value 0.33. 
Table~\ref{ge} shows the genre node(s), number of movies and users (percentage of the total) contained in each community. 
We can see Community 2 is the most popular community among users (contains about 36\% of total users) and Community 7 is the least popular community among users (contains less than 2\% of the total users).
An interesting observation is that each community contains a distinctive type of movies. 
This observation can in turn help us understand the movie preferences of users contained in each community.
For example, users in Community 6 like movies from the animation, children's, fantasy and musical genre. This preference is very different from that of users in Community 2 who like movies from the crime, file-noir, mystery and thriller genre. 
We can also see horror and documentary each form its own small community.

\begin{table}[!htb]
\centering
\begin{tabular}{c|l|l|l}
\hline
Community & \multicolumn{1}{c|}{Movie Genre}        & \begin{tabular}[c]{@{}l@{}}\# of movies\\ (\% of total)\end{tabular} & \begin{tabular}[c]{@{}l@{}}\# of users\\ (\% of total)\end{tabular} \\ \hline
1         & Drama, War                              & 27\%                                                                         & 21\%                                                                        \\
2         & Crime, Film-Noir, Mystery, Thriller     & 16\%                                                                         & 36\%                                                                        \\
3         & Horror                                  & 5\%                                                                          & 3\%                                                                         \\
4         & Action, Adventure, Sci-Fi, Western      & 14\%                                                                         & 17\%                                                                        \\
5         & Comedy, Romance                         & 26\%                                                                         & 16\%                                                                        \\
6         & Animation, Children's, Fantasy, Muscial & 9\%                                                                          & 6\%                                                                         \\
7         & Documentary                             & 3\%                                                                          & 1\%                                                                         \\ \hline
\end{tabular}
\caption{Movie genre, number of movies and users contained in the identified communities.}
\label{ge}
\end{table}

In the MovieLens dataset, demographic information such as gender and occupation is available for some of the users. Over 70\% of the identified male users are in Communities 1, 2 and 4; over 60\% of the identified female users are in Communities 1 and 2. 
We find that Communities 3 and 7 are the least popular among users who are listed as students (the two communities together contain less than 4\% of the student users). 
Communities 1 and 2 are the most popular among users who are listed as educators or administrators (the two communities together contain over 70\% of the educator users and 55\% of the administrator users). 
Community 4 is the most popular among users who are listed as programmers or engineers (this community contains over 30\% of the engineer users and over 30\% of the programmer users).

\vspace{3pt}
\setcounter{chapter}{7}
\setcounter{equation}{0}
\noindent {\bf 7. Discussion}
\vspace{3pt}

In this paper, we propose a modularity based framework for community detection on heterogeneous networks. 
Specifically, we define a null model for heterogeneous networks. 
Furthermore, we propose a modularity maximization method that can handle very large networks. We show that under a heterogeneous stochastic blockmodel, the proposed modularity function is consistent as a community detection criterion.
The proposed community detection approach performs well with both simulated and real-world networks.

Since the modularity maximization problem is NP-hard, existing approaches are heuristic approaches, and are not guaranteed to find the global maximizer of the function. Even though the Louvain method shows good performance in our simulation studies and other reported studies, there lacks a thorough investigation of its theoretical property. This is an important problem that worth investigating next.

We note that the maximization of the proposed modularity function is not tied to the Louvain method. In fact, several existing modularity maximization techniques can be applied to our setting with some modifications, such as the spectral method based on the eigen decomposition of the modularity matrix (Newman, 2006) or the stochastic maximization method (Massen and Doye, 2005). However, in practice, we find that the Louvain method yields better modularity maximum than the other methods and is computationally more efficient. 
Another suitable approach is to apply the spectral method proposed in Sengupta and Chen (2015), which performs a K-means clustering of the $K$ eigenvectors corresponding to the $K$ largest eigenvalues of the regularized graph Laplacian matrix. 

The proposed method can be extended to directed heterogeneous networks. Several approaches have been proposed for finding communities in directed homogeneous networks using modified modularity functions (see Fortunato, 2010 for a review). To incorporate directed edges into our framework, we need to define a null model for directed heterogeneous networks. Furthermore, we need to calculate the expectations under the null model. This is an interesting topic to investigate.

\vskip 14pt
\noindent{\large\bf Supplementary Material}

The online supplementary material includes proofs for Theorem 1 and Theorem 2.
\par

\vspace{10pt}
\noindent {\bf References}

\begin{description}\setlength{\itemsep}{-2ex}

\item Agrawal, G., and Kempe, D. (2008). Modularity-maximizing graph communities via mathematical programming. \textit{The European Physics Journal B} \textbf{66}, 409-418.

\item Airoldi, E. M., Blei, D. M., Fienberg, S. E., and Xing, E. P. (2008). Mixed-membership stochastic blockmodels. \textit{Journal of Machine Learning Research} \textbf{9}, 1981-2014.

\item Bender, E. and Canfield, R. (1978), ``The asymptotic number of labeled graphs with given degree sequences," \textit{Journal of Combinatorial Theory A} \textbf{24}, 296-307.

\item Bickel, P., and Chen, A. (2009). A non-parametric view of network models and Newman-Girvan and other modularities. \textit{Proceedings of the National Academy of Sciences} \textbf{106}, 21068-21073.

\item Blondel, V. D., Guillaume, J. L., Lambiotte, R., and Lefebvre, E. (2008). Fast unfolding of communities in large networks. \textit{Journal of Statistical Mechanics: Theory and Experiment} \textbf{10}, P10008.

\item Blondel, V. D. (2011). The Louvain method for community detection in large networks. \\
\textit{https://perso.uclouvain.be/vincent.blondel/research/louvain.html}.

\item Bollob\'{a}s, B., and McKay, B. D. (1986). The number of matchings in random regular graphs and bipartite graphs. \textit{Journal of Combinatorial Theory, Series B} \textbf{41(1)}, 80-91.

\item Brandes, U., Delling, D., Gaertler, M., Gorke, R., Hoefer, M., Nikoloski, Z., and Wagner, D. (2008). On modularity clustering. \textit{IEEE Transactions on Knowledge and Data Engineering} \textbf{20}, 172-188.

\item Chung, F., and Lu, L. (2002). Connected components in random graphs with given expected degree sequences. \textit{Annals of Combinatorics} \textbf{6(2)}, 125-145.

\item Clauset, A., Newman, M. E. J., and Moore, C. (2004). Finding community structure in very large networks. \textit{Physical Review E} \textbf{70}, 066111.

\item Danon, L., Diaz-Guilera, A., Duch, J., and Arenas, A. (2005). Comparing community structure identification. \textit{Journal of Statistical Mechanics: Theory and Experiment} \textbf{09}, P09008.

\item Fortunato, S. (2010). Community detection in graphs. \textit{Physics Reports} \textbf{428}, 75-174.

\item Gao, J., Liang, F., Fan, W., Sun, Y., and Han, J. (2009). Graph-based consensus maximization among multiple supervised and unsupervised models. \textit{Advances in Neural Information Processing Systems} \textbf{22}, 585-593.

\item Harper, F. M., and Konstan, J. A. (2016). The movielens datasets: history and context. \textit{ACM Transactions on Interactive Intelligent Systems} \textbf{5(4)}, 19.

\item Massen, C., and Doye, J. (2005). Identifying communities within energy landscapes. \textit{Physical Review E} \textbf{71}, 046101.


\item Liu, X., Liu, W., Murata, T., and Wakita, K. (2014). A framework for community detection in heterogeneous multi-relational networks. \textit{Advances in Complex Systems} \textbf{17(06)}, 1450018.

\item Ji, M., Sun, Y., Danilevsky, M., Han, J., and Gao, J. (2010). Graph regularized transductive classification on heterogeneous information networks. \textit{Proceeding of the Joint European Conference on Machine Learning and Knowledge Discovery in Databases}, 570-586.

\item Jin, J. (2015). Fast community detection by SCORE. \textit{The Annals of Statistics} \textbf{43}, 57-89.

\item Newman, M. E. J. (2006). Finding community structure in networks using the eigenvectors of matrices. \textit{Physical Review E} \textbf{74}, 035104.

\item Newman, M. E. J., and Girvan, M. (2004). Finding and evaluating community structure in networks. \textit{Physical Review E} \textbf{69}, 026113.

\item Perry, P., and Wolfe, P. (2012). Null models for network data. https://arxiv.org/abs/1201.5871.

\item Reichardt, J., and Bornholdt, S. (2006). Statistical mechanics of community detection. \textit{Physical Review E} \textbf{74}, 016110.

\item Rohe, K., Chatterjee, S., and Yu, B. (2011). Spectral clustering and the high-dimensional stochastic blockmodel. \textit{The Annals of Statistics} \textbf{39(4)}, 1878-1915.

\item Sengupta, S., and Chen, Y. (2015). Spectral clustering in heterogeneous networks. \textit{Statistica Sinica} \textbf{25}, 1081-1106.

\item Shi, J., and Malik, J. (2000). Normalized cuts and image segmentation. \textit{IEEE Transactions on Pattern Analysis and Machine Intelligence} \textbf{22}, 888-905.

\item Sun, Y., and Han, J. (2012). Mining heterogeneous information networks: principles and methodologies. \textit{Synthesis Lectures on Data Mining and Knowledge Discovery} \textbf{3(2)}, 1-159.

\item Wakita, K., and Tsurumi, T. (2007). Finding community structure in mega-scale social networks. \textit{Proceedings of the 16th International Conference on World Wide Web}, 1275-1276.

\item Zhang, J., and Cao, J. (2017). Finding common modules in a time-varying network with application to the Drosophila Melanogaster gene regulation network. \textit{Journal of the American Statistical Association}, in press.

\item Zhang, J., and Chen, Y. (2016). A hypothesis testing framework for modularity based network community detection. \textit{Statistica Sinica} \textbf{27}, 437-456.

\item Zhao, Y., Levina, E., and Zhu, J. (2012). Consistency of community detection in networks under degree-corrected stochastic block models. \textit{The Annals of Statistics} \textbf{40}, 2266-2292.

\end{description}

\vskip .65cm
\noindent
Department of Management Science, University of Miami, Coral Gables, FL 33124, USA.
\vskip 2pt
\noindent
E-mail: ezhang@bus.miami.edu
\vskip 2pt

\noindent
Department of Statistics, University of Illinois at Urbana-Champaign, Champaign, IL 61820, USA.
\vskip 2pt
\noindent
E-mail: yuguo@illinois.edu

\vspace{.55cm}
 \centerline{\bf Supplementary Material}
\vspace{.55cm}

\def\theequation{S\arabic{section}.\arabic{equation}}
\def\thesection{S\arabic{section}}
\def\thetheorem{S\arabic{theorem}}

\section{Proof of Theorem 1}
First, we state a theorem from Bollob\'{a}s and McKay (1986) on the asymptotic number of simple graphs with forbidden edges. 
Consider simple graphs with $m$ edges and degree sequence $\bm{d}=(d_1, \ldots,d_n)$. 
Let $D$ be an $n\times n$ symmetric zero-one matrix that specifies the set of forbidden edges ($D_{ij}=D_{ji}=1$ if an edge between node $i$ and node $j$ is forbidden, and $D_{ij}=D_{ji}=0$ otherwise).
Define $\lambda=\sum_{i=1}^n d_i(d_i-1)/(4m)$, $\gamma=\sum_{{i<j, D_{ij}=1}}\frac{d_id_j}{2m}$,
$d_{max}=\max_i d_i$. We have the following theorem.
\begin{theorem} (Bollob\'{a}s and McKay, 1986)
\label{thms1}
Suppose the maximum column (row) sum of $D$ is at most $(\log n)^{1/3}$, $d_{max}\le(\log n)^{1/3}$, and for some $\eta>0$, $m\ge\max\{\eta d_{max},(1+\eta)n\}$. Then the number of simple graphs with degree sequence $\bm{d}$ and none of the forbidden edges specified in $D$ is 
\begin{equation}
(1+\epsilon) \exp\{-\lambda-\lambda^2-\gamma\}\frac{(2m)!}{(m!)2^{m}\prod_{i=1}^n d_i!},
\label{mckay_approx}
\end{equation}
where $\epsilon\le o(e^{(\log n)^{4/5}}/n)$.
\end{theorem}

From the development in Section 2.2, we have
\begin{eqnarray*}
E(A_{ij}^{[l]})&=&\frac{|\Sigma_{\mathbf{d}^{[l]}|{A_{ij}^{[l]}=1}}|}{|\Sigma_{\mathbf{d}^{[l]}}|}=1-\frac{|\Sigma_{\mathbf{d}^{[l]}|{A_{ij}^{[l]}=0}}|}{|\Sigma_{\mathbf{d}^{[l]}}|}, \quad l=1,2,
\end{eqnarray*}
where $|\Sigma_{\mathbf{d}^{[l]}|{A_{ij}^{[l]}=0}}|$ is the total number of simple homogeneous networks with degree sequence $\mathbf{d}^{[l]}$ and no link between nodes $i$ and $j$.
Using (\ref{mckay_approx}), we can approximate $|\Sigma_{\mathbf{d}^{[l]}|{A_{ij}^{[l]}=1}}|$ and $|\Sigma_{\mathbf{d}^{[l]}}|$, which will lead to the approximation for $E(A_{ij}^{[l]})$, $l=1,2$.

For the set $|\Sigma_{\mathbf{d}^{[l]}}|$, $l=1,2$, the matrix $D$ has 0 for all entries. Directly applying Theorem~\ref{thms1}, we have 
\begin{equation}
|\Sigma_{\mathbf{d}^{[l]}}|=(1+\epsilon_l) \exp\{-\lambda^{[l]}-{\lambda^{[l]}}^2\}\frac{(2m^{[l]})!}{(m^{[l]}!)2^{m^{[l]}}\prod_{i=1}^n d^{[l]}_i!},
\label{e1}
\end{equation}
where $\lambda^{[l]}=\sum_{i=1}^n d^{[l]}_i(d^{[l]}_i-1)/(4m^{[l]})$, and $\epsilon_l\le o(e^{(\log n_l)^{4/5}}/n_l)$, $l=1,2$.
For the set $|\Sigma_{\mathbf{d}^{[l]}|{A_{ij}^{[l]}=0}}|$, $l=1,2$, the matrix $D$ has $D_{ij}=D_{ji}=1$ and 0 elsewhere. 
From this, we immediately have that the maximum column (row) sum of $D$ is less than $(\log n_l)^{1/3}$ and $\gamma=d^{[l]}_id^{[l]}_j/(2m^{[l]})$.
Again directly applying Theorem~\ref{thms1}, we have 
\begin{equation}
|\Sigma_{\mathbf{d}^{[l]}|{A_{ij}^{[l]}=0}}|=(1+\epsilon_l^*) \exp\{-\lambda^{[l]}-{\lambda^{[l]}}^2-\frac{d^{[l]}_id^{[l]}_j}{2m^{[l]}}\}\frac{(2m^{[l]})!}{(m^{[l]}!)2^{m^{[l]}}\prod_{i=1}^n d^{[l]}_i!},
\label{e2}
\end{equation}
where $\epsilon^*_l\le o(e^{(\log n_l)^{4/5}}/n_l)$, $l=1,2$.
From (\ref{e1}) and (\ref{e2}), we have
\begin{eqnarray*}
\frac{|\Sigma_{\mathbf{d}^{[l]}|{A_{ij}^{[l]}=0}}|}{|\Sigma_{\mathbf{d}^{[l]}}|}&=&(1+o(e^{(\log n_l)^{4/5}}/n_l))e^{-\frac{d^{[l]}_id^{[l]}_j}{2m^{[l]}}}\\
&=&e^{-\frac{d^{[l]}_id^{[l]}_j}{2m^{[l]}}}+o(e^{(\log n_l)^{4/5}}/n_l),
\end{eqnarray*}
where the second equality is true because $e^{-\frac{d^{[l]}_id^{[l]}_j}{2m^{[l]}}}\le e^0=1$.
Therefore, we have as $n_l\rightarrow\infty$
\begin{eqnarray*}
E(A_{ij}^{[l]})&=&1-e^{-\frac{d^{[l]}_id^{[l]}_j}{2m^{[l]}}}+o(e^{(\log n_l)^{4/5}}/n_l)\\
&=&\frac{d^{[l]}_id^{[l]}_j}{2m^{[l]}}+o\left(\frac{d^{[l]}_id^{[l]}_j}{2m^{[l]}}\right)+o(e^{(\log n_l)^{4/5}}/n_l)\\
&=&\frac{d^{[l]}_id^{[l]}_j}{2m^{[l]}}+o(e^{(\log n_l)^{4/5}}/n_l),
\end{eqnarray*}
where the last equality is true because $\frac{d^{[l]}_id^{[l]}_j}{2m^{[l]}}\le (\log n_l)^{2/3}/(2n_l(1+\eta))$, and $(\log n_l)^{2/3}/n_l$ converges to 0 faster than $e^{(\log n_l)^{4/5}}/n_l$ as $n_l\rightarrow\infty$, $l=1,2$.

Next it remains for us to show that as $n_1,n_2\rightarrow\infty$,
$$
E(A_{ij}^{[12]})=\frac{d_i^{[12]}d_j^{[21]}}{m^{[12]}}+o(n_2^{-3/4}).
$$
First, it is easy to derive that 
\begin{equation*}
E(A_{ij}^{[12]})=\frac{|\Sigma_{\mathbf{d}^{[12]},\mathbf{d}^{[21]}|{A_{ij}^{[12]}=1}}|}{|\Sigma_{\mathbf{d}^{[12]},\mathbf{d}^{[21]}}|}=1-\frac{|\Sigma_{\mathbf{d}^{[12]},\mathbf{d}^{[21]}|{A_{ij}^{[12]}=0}}|}{|\Sigma_{\mathbf{d}^{[12]},\mathbf{d}^{[21]}}|},
\end{equation*}
where $|\Sigma_{\mathbf{d}^{[12]},\mathbf{d}^{[21]}|{A_{ij}^{[12]}=0}}|$ is the total number of bi-partite graphs with degree sequences $\mathbf{d}^{[12]}$ for type-$[1]$ nodes, $\mathbf{d}^{[21]}$ for type-$[2]$ nodes and no link between the $i$th node of type-$[1]$ and the $j$th node of type-$[2]$.
We will find asymptotic formulae for  $|\Sigma_{\mathbf{d}^{[12]},\mathbf{d}^{[21]}|{A_{ij}^{[12]}=0}}|$ and $|\Sigma_{\mathbf{d}^{[12]},\mathbf{d}^{[21]}}|$, which in turn lead to the approximation for $E(A_{ij}^{[12]})$.

Next, we state the following theorem from Bollob\'{a}s and McKay (1986). It is an analog of Theorem~\ref{thms1} for bipartite graphs. 
Consider simple bipartite graphs with $m$ edges and degree sequence $\bm{d}=(d_1, \ldots,d_{n_1})$ for one type of nodes, referred to as type-[1] nodes, and $\bm{d'}=(d'_1, \ldots,d'_{n_2})$ for the other type of nodes, referred to type-[2] nodes. Suppose $n_1\le n_2$.
Let $D$ be an $n_1\times n_2$ zero-one matrix that specifies the set of forbidden edges ($D_{ij}=1$ if an edge between node $i$ of type-[1] and node $j$ of type-[2] is forbidden, and $D_{ij}=0$ otherwise).
Define $\lambda=(\sum_{i=1}^{n_1} d_i(d_i-1))(\sum_{i=1}^{n_2} d'_i(d'_i-1))/m^2$, $\gamma=\sum_{{i,j, D_{ij}=1}}\frac{d_id_j}{m}$,
$d_{max}=\max_i d_i$ and $d'_{max}=\max_i d'_i$. We have the following theorem.
\begin{theorem} (Bollob\'{a}s and McKay, 1986)
\label{thms2}
Suppose the maximum column and row sums of $D$ are at most $(\log n_1)^{1/3}$, $d_{max}\le(\log n_1)^{1/3}$, $d'_{max}\le(\log n_1)^{1/3}$, and for some $\eta>0$, $m\ge\max\{\eta d_{max},\eta d'_{max},(2+\eta)n_2\}$. Then the number of bipartite graphs with degree sequence $\bm{d}$, $\bm{d'}$ and none of the forbidden edges specified in $D$ is 
\begin{equation}
(1+\epsilon) \exp\{-\lambda-\gamma\}m!/\left\{\prod_{i=1}^{n_1} d_i!\prod_{i=1}^{n_2} d'_i!\right\},
\end{equation}
where $\epsilon=o(n_2^{-3/4})$.
\end{theorem}
From here, the arguments used to derive $E(A_{ij}^{[12]})$ is similar to those used in derive $E(A_{ij}^{[l]})$, $l=1,2$. Here we omit the details.
 
\section{Proof of Theorem 2}
First we formalize the notations that will be used in the proof. Consider a heterogeneous network $\mathcal{G}(\bigcup_{i=1}^LV^{[i]},\mathcal{E}\cup\mathcal{E}^+)$. For a community assignment label $\mathcal{E}=(\be^{[1]},\ldots,\be^{[L]})$ with $\be^{[l]}=(e_1^{[l]}, \ldots,e_{n_l}^{[l]})$, $l=1,\dots,L$, define $K\times K$ matrices $O^{[l]}$, $l=1,\dots,L$, and $O^{[l_1l_2]}$, $1\le l_1\neq l_2\le L$, such that
$$
O^{[l]}_{kh}(\mathcal{E})=\sum_{ij}A_{ij}^{[l]}I(e^{[l]}_i=k,e^{[l]}_j=h),
$$
$$
O^{[l_1l_2]}_{kh}(\mathcal{E})=\sum_{ij}A_{ij}^{[l_1l_2]}I(e^{[l_1]}_i=k,e^{[l_2]}_j=h).
$$
Define $O^{[l]}_{k}=\sum_hO^{[l]}_{kh}$ and $O^{[l_1l_2]}_{k}=\sum_hO^{[l_1l_2]}_{kh}$, $l=1,\dots,L$, $1\le l_1\neq l_2\le L$. 
Define $K\times K$ matrices $R^{[l]}(\mathcal{E})$, $V^{[l]}(\mathcal{E})$, $l=1,\ldots,L$, such that
$$
R^{[l]}_{ab}(\mathcal{E})=\frac{1}{n}\sum_{l=1}^{n_l}I(e^{[l]}_i=a,c^{[l]}_i=b)
$$
$$
V_{ab}^{[l]}(\mathcal{E})=\frac{\sum_{l=1}^{n_l}I(e^{[l]}_i=a,c^{[l]}_i=b)}{\sum_{l=1}^{n_l}I(c^{[l]}_i=b)}.
$$
Write $\mathcal{O}=\{O^{[l]}, O^{[l_1l_2]}, l=1,\ldots,L, 1\le l_1\neq l_2\le L\}$ and $\mathcal{R}=\{R^{[1]},\ldots,R^{[L]}\}$. 

For community assignment label $\mathcal{E}$, the contribution of the bipartite graph $G_{l_1l_2}$ to the modularity function $Q'(\mathcal{E},\mathcal{G})$ is 
$$
q_{l_1l_2}=\frac{1}{L^2}\sum_{ij}\left(A_{ij}^{[l_1l_2]}-\frac{d_i^{[l_1l_2]}d_j^{[l_2l_1]}}{m^{[l_1l_2]}}\right)\delta(e^{[l_1]}_i,e^{[l_2]}_j),
$$
where $\delta(\cdot,\cdot)$ is the Kronecker function. We have
\begin{eqnarray*}
q_{l_1l_2}&=&\frac{1}{L^2}\left(\sum_{ij}A_{ij}^{[l_1l_2]}\delta(e^{[l_1]}_i,e^{[l_2]}_j)-\frac{1}{m^{[l_1l_2]}}\sum_k\sum_{ij}d_i^{[l_1l_2]}d_j^{[l_2l_1]}I(e_i^{[l_1]}=k)I(e^{[l_2]}_j=k))\right)\\
&=&\frac{1}{L^2}\left(\sum_kO_k^{[l_1l_2]}-\frac{1}{m^{[l_1l_2]}}\sum_kO_k^{[l_1l_2]}O_k^{[l_2l_1]}\right).
\end{eqnarray*}
Following similar arguments, it is easy to show that the modularity function $Q'(\mathcal{E},\mathcal{G})$ can be expressed as 
$$
\frac{1}{L^2}\left[\sum_{l=1}^L\sum_{k=1}^K\left(O^{[l]}_{kk}-\frac{{O^{[l]}_k}^2}{\sum_{kh}O^{[l]}_{kh}}\right)+\sum_{l_1\neq l_2}^L\sum_{k=1}^K\left(O^{[l_1l_2]}_{kk}-\frac{O^{[l_1l_2]}_kO^{[l_2l_1]}_k}{\sum_{kh}O^{[l_1l_2]}_{kh}}\right)\right].
$$ 
Here we suppress the argument $\mathcal{E}$ for brevity. 
Define 
$$
J(\mathcal{O})=\sum_{l=1}^LJ_1(O^{[l]})+\sum_{l_1\neq l_2}^LJ_2(O^{[l_1l_2]},O^{[l_2l_1]}),
$$
where 
$$
J_1(O^{[l]})=\sum_{k=1}^K\left(O^{[l]}_{kk}-\frac{{O^{[l]}_k}^2}{\sum_{kh}O^{[l]}_{kh}}\right),
$$
and 
$$
J_2(O^{[l_1l_2]},O^{[l_2l_1]})=\sum_{k=1}^K\left(O^{[l_1l_2]}_{kk}-\frac{O^{[l_1l_2]}_kO^{[l_2l_1]}_k}{\sum_{kh}O^{[l_1l_2]}_{kh}}\right).
$$
We show the consistency property by showing that there exists $\delta_n\rightarrow 0$ such that
$$
P\left(\max_{\mathcal{E}:\ \eta(\mathcal{E},\mathcal{C})\ge\delta_n}J\left(\frac{\mathcal{O}(\mathcal{E})}{\mu_n}\right)\le J\left(\frac{\mathcal{O}(\mathcal{C})}{\mu_n}\right)\right)\rightarrow1\quad\text{as}\quad n\rightarrow\infty,
$$
where $\eta(\mathcal{E},\mathcal{C})=\sum_{l=1}^L\sum_{ab}|V_{ab}^{[l]}(\mathcal{E})-V_{ab}^{[l]}(\mathcal{C})|$.

Define $\mu_n=n^2\rho_n$, we have
\begin{eqnarray*}
&&\frac{1}{\mu_n}E(O^{[l_1l_2]}_{kh}(\mathcal{E})\mid\mathcal{C})\\
&&=\frac{1}{\mu_n}E\left(\sum_{ij}A_{ij}^{[l_1l_2]}I(e_i^{[l_1]}=k,e_j^{[l_2]}=h)\mid\mathcal{C}\right)\\
&&=\frac{1}{n^2}\sum_{ij}\sum_{ab}P_{ab}^{[l_1l_2]}I(e_i^{[l_1]}=k,c_i^{[l_1]}=a)I(e_j^{[l_2]}=h,e_j^{[l_2]}=b).
\end{eqnarray*}
Define $H^{[l_1l_2]}(\mathcal{R}(\mathcal{E}))=\frac{1}{\mu_n}E(O^{[l_1l_2]}(\mathcal{E})\mid\mathcal{C})$, we have
$$
H^{[l_1l_2]}(\mathcal{R}(\mathcal{E}))=R^{[l_1]}(\mathcal{E})P^{[l_1l_2]}R^{[l_2]}(\mathcal{E})',\quad 1\le l_1\neq l_2\le L.
$$
Similarly, we can define $H^{[l]}(\mathcal{R}(\mathcal{E}))=\frac{1}{\mu_n}E(O^{[l]}(\mathcal{E})\mid\mathcal{C})$ and write
$$
H^{[l]}(\mathcal{R}(\mathcal{E}))=R^{[l]}(\mathcal{E})P^{[l]}R^{[l]}(\mathcal{E})',\quad l=1,\ldots,L.
$$
Write $\mathcal{H}=\{H^{[l]}, H^{[l_1l_2]}, l=1,\dots,L, 1\le l_1\neq l_2\le L\}$. Since $J(.)$ is Lipschitz in all its arguments, we have 
\begin{eqnarray*}
\left|J\left(\frac{\mathcal{O}(\mathcal{E})}{\mu_n}\right)-J\left(\mathcal{H}(\mathcal{R})\right)\right|\le M_1\left[\max_{l}\parallel\frac{O^{[l]}(\mathcal{E})}{\mu_n}-H^{[l]}(\mathcal{R})\parallel_{\infty}+\max_{l_1\neq l_2}\parallel\frac{O^{[l_1l_2]}(\mathcal{E})}{\mu_n}-H^{[l_1l_2]}(\mathcal{R})\parallel_{\infty}\right].
\end{eqnarray*}
Here $||X||_{\infty}=\max_{kh}|X_{kh}|$.
To continue with the proof, we need to use the Bernstein's inequality (Bernstein, 1924).

\noindent\textbf {Bernstein's inequality:} \textit{Let $X_1,\ldots,X_n$ be independent variables. Suppose that $|X_i|\le M$ for all $i$. Then, for all positive t,
\begin{equation*}
P\left(\left|\sum_{i=1}^nX_i-\sum_{i=1}^nE(X_i)\right|>t\right)\le2\exp\left(-\frac{t^2/2}{\sum \text{var}(X_i)+Mt/3}\right).
\end{equation*}}
Since $A^{[l]}_{ij}$'s in $O^{[l]}(\mathcal{E})$ are independent Bernoulli random variables, applying the Bernstein's inequality, we have 
\begin{equation*}
P\left(|O^{[l]}_{kh}(\mathcal{E})/\mu_n-H^{[l]}_{kh}(\mathcal{R})|>\omega\right)\le 2\exp\left(-\frac{\omega^2/2}{\text{var}(O^{[l]}_{kh}(\mathcal{E}))+2\omega/3}\right).
\end{equation*}
Notice that $\text{var}(O^{[l]}_{kh}(\mathcal{E}))\le 2n^2\max_{ij}\text{var}(A^{[l]}_{ij})$.

Define $\tau=\max_{ij}\text{var}(A^{[l]}_{ij})$. For any $\epsilon<3\tau$, if we write $\omega=\epsilon n^2\rho_n$, then we have
\begin{eqnarray*}
P\left(\left|\frac{O^{[l]}_{kh}(\mathcal{E})}{\mu_n}-H^{[l]}_{kh}(\mathcal{R})\right|>\epsilon\right) &\le& 2\exp\left(-\frac{\omega^2/2}{\text{var}(O^{[l]}_{kh}(\mathcal{E}))+2\omega/3}\right)\\
&\le& 2\exp\left(-\frac{\epsilon^2n^4\rho_n^2}{8n^2\rho_n\tau}\right)\\
&=& 2\exp\left(-\frac{\epsilon^2\mu_n}{8\tau}\right).
\end{eqnarray*}
The left hand side of the inequality converges to 0 in probability uniformly over $\mathcal{E}$ as $\lambda_n\rightarrow\infty$. Following similar arguments, we can show that 
$$
P\left(\left|\frac{O^{[l_1l_2]}_{kh}(\mathcal{E})}{\mu_n}-H^{[l_1l_2]}_{kh}(\mathcal{R})\right|>\epsilon\right)
$$
also converges to 0 in probability uniformly as $\lambda_n\rightarrow\infty$.
Thus, there exists $\epsilon_n\rightarrow0$, such that
\begin{equation}
\label{pop}
P\left(\max_{\mathcal{E}}\left|J\left(\frac{\mathcal{O}(\mathcal{E})}{\mu_n}\right)-J\left(\mathcal{H}(\mathcal{R})\right)\right|\le\epsilon_n\right)\rightarrow1\quad\text{as}\quad\lambda_n\rightarrow\infty.
\end{equation}

Next we show that $J\left(\mathcal{H}(\mathcal{R})\right)$ is uniquely maximized over $\{\mathcal{R}: R^{[l]}\ge0,\ {R^{[l]}}'\textbf{1}=\pi^{[l]},\ l=1,\ldots,L\}$ at $\mathcal{S}=\mathcal{R}(\mathcal{C})$. Since $J\left(\mathcal{H}(\mathcal{R})\right)$ is the population version of $J\left(\frac{\mathcal{O}(\mathcal{E})}{\mu_n}\right)$, if $J\left(\frac{\mathcal{O}(\mathcal{E})}{\mu_n}\right)$ is maximized by the true community label $\mathcal{C}$, $J\left(\mathcal{H}(\mathcal{R})\right)$ should also be maximized by the true assignment $\mathcal{S}$.

Define
\begin{equation*}
\bigtriangleup_{kh}=\left\{\begin{array}{rcl}
1& \mbox{for}& k=h, \\
-1& \mbox{for} & k\neq h.
\end{array}\right.
\end{equation*}

Using the equalities
$$
\sum_k\left(H^{[l]}_{kk}-\frac{{H^{[l]}_k}^2}{\sum_{kh}H^{[l]}_{kh}}\right)+\sum_{k\neq l}\left(H^{[l]}_{kh}-\frac{H^{[l]}_kH^{[l]}_h}{\sum_{kh}H^{[l]}_{kh}}\right)=0,\quad l=1,\ldots,L,
$$
and
$$
\sum_k\left(H^{[l_1l_2]}_{kk}-\frac{H^{[l_1l_2]}_kH^{[l_2l_1]}_k}{\sum_{kh}H^{[l_1l_2]}_{kh}}\right)+\sum_{k\neq h}\left(H^{[l_1l_2]}_{kh}-\frac{H^{[l_1l_2]}_kH^{[l_2l_1]}_h}{\sum_{kh}H^{[l_1l_2]}_{kh}}\right)=0,\quad1\le l_1\neq l_2\le L.
$$
we have 
\begin{eqnarray}
\nonumber
&&J(\mathcal{H}(\mathcal{R}))=\sum_{l=1}^LJ_1(H^{[l]}(\mathcal{R}))+\sum_{l_1\neq l_2}^LJ_2(H^{[l_1l_2]}(\mathcal{R}),H^{[l_2l_1]}(\mathcal{R}))\\\nonumber
&=&\frac{1}{2}\sum_{l=1}^L\sum_{kh}\bigtriangleup_{kh}\left(H^{[l]}_{kh}(\mathcal{R})-\frac{H^{[l]}_k(\mathcal{R})H^{[l]}_h(\mathcal{R})}{\sum_{kh}H^{[l]}_{kh}(\mathcal{R})}\right)+\frac{1}{2}\sum_{l_1\neq l_2}^L\sum_{kh}\bigtriangleup_{kh}\left(H^{[l_1l_2]}_{kh}(\mathcal{R})-\frac{H^{[l_1l_2]}_k(\mathcal{R})H^{[l_2l_1]}_h(\mathcal{R})}{\sum_{kh}H^{[l_1l_2]}_{kh}(\mathcal{R})}\right)\\\nonumber
&=&\frac{1}{2}\sum_{l=1}^L\sum_{kh}\bigtriangleup_{kh}\left(\sum_{ab}P^{[l]}_{ab}R^{[l]}_{ka}(\mathcal{E})R^{[l]}_{hb}(\mathcal{E})-\frac{(\sum_{as}P^{[l]}_{as}R^{[l]}_{ka}(\mathcal{E})\pi^{[l]}_{s})(\sum_{bt}P^{[l]}_{bt}R^{[l]}_{hb}(\mathcal{E})\pi^{[l]}_{t})}{\sum_{kh}H^{[l]}_{kh}(\mathcal{R})}\right)\\\nonumber
&&+\frac{1}{2}\sum_{l_1\neq l_2}^L\sum_{kh}\bigtriangleup_{kh}\left(\sum_{ab}P^{[l_1l_2]}_{ab}R^{[l_1]}_{ka}(\mathcal{E})R^{[l_2]}_{hb}(\mathcal{E})-\frac{(\sum_{as}P^{[l_1l_2]}_{as}R^{[l_1]}_{ka}(\mathcal{E})\pi^{[l_2]}_{s})(\sum_{bt}P^{[l_2l_1]}_{bt}R^{[l_2]}_{hb}(\mathcal{E})\pi^{[l_1]}_{t})}{\sum_{kh}H^{[l_1l_2]}_{kh}(\mathcal{R})}\right)\\\nonumber
&=&\frac{1}{2}\sum_{l=1}^L\sum_{kh}\sum_{ab}\bigtriangleup_{kh}R^{[l]}_{ka}(\mathcal{E})R^{[l]}_{hb}(\mathcal{E})\left(P^{[l]}_{ab}-\frac{(\sum_{s}P^{[l]}_{as}\pi^{[l]}_{s})(\sum_{t}P^{[l]}_{bt}\pi^{[l]}_{t})}{\sum_{kh}H^{[l]}_{kh}(\mathcal{R})}\right)\\\nonumber
&&+\frac{1}{2}\sum_{l_1\neq l_2}^L\sum_{kh}\sum_{ab}\bigtriangleup_{kh}R^{[l_1]}_{ka}(\mathcal{E})R^{[l_2]}_{hb}(\mathcal{E})\left(P^{[l_1l_2]}_{ab}-\frac{(\sum_{s}P^{[l_1l_2]}_{as}\pi^{[l_2]}_{s})(\sum_{t}P^{[l_2l_1]}_{bt}\pi^{[l_1]}_{t})}{\sum_{kh}H^{[l_1l_2]}_{kh}(\mathcal{R})}\right)\\\nonumber
\end{eqnarray}
\begin{eqnarray}
\nonumber
&\le&\frac{1}{2}\sum_{l=1}^L\sum_{kh}\sum_{ab}\bigtriangleup_{ab}R^{[l]}_{ka}(\mathcal{E})R^{[l]}_{hb}(\mathcal{E})\left(P^{[l]}_{ab}-\frac{(\sum_{s}P^{[l]}_{as}\pi^{[l]}_{s})(\sum_{t}P^{[l]}_{bt}\pi^{[l]}_{t})}{\sum_{kh}H^{[l]}_{kh}(\mathcal{R})}\right)\\\nonumber
&&+\frac{1}{2}\sum_{l_1\neq l_2}^L\sum_{kh}\sum_{ab}\bigtriangleup_{ab}R^{[l_1]}_{ka}(\mathcal{E})R^{[l_2]}_{hb}(\mathcal{E})\left(P^{[l_1l_2]}_{ab}-\frac{(\sum_{s}P^{[l_1l_2]}_{as}\pi^{[l_2]}_{s})(\sum_{t}P^{[l_1l_2]}_{bt}\pi^{[l_1]}_{t})}{\sum_{kh}H^{[l_1l_2]}_{kh}(\mathcal{R})}\right)\\\nonumber
&=&\frac{1}{2}\sum_{l=1}^L\sum_{ab}\bigtriangleup_{ab}\pi^{[l]}_{a}\pi^{[l]}_{b}\left(P^{[l]}_{ab}-\frac{(\sum_{s}P^{[l]}_{as}\pi^{[l]}_{s})(\sum_{t}P^{[l]}_{bt}\pi^{[l]}_{t})}{\sum_{kh}H^{[l]}_{kh}(\mathcal{S})}\right)\\\nonumber
&&+\frac{1}{2}\sum_{l_1\neq l_2}^L\sum_{ab}\bigtriangleup_{ab}\pi^{[l_1]}_{a}\pi^{[l_2]}_{b}\left(P^{[l_1l_2]}_{ab}-\frac{(\sum_{s}P^{[l_2l_1]}_{as}\pi^{[l_2]}_{s})(\sum_{t}P^{[l_1l_2]}_{bt}\pi^{[l_1]}_{t})}{\sum_{kh}H^{[l_1l_2]}_{kh}(\mathcal{S})}\right)\\\nonumber
&=&\sum_{l=1}^LJ_1(H^{[l]}(\mathcal{S}))+\sum_{l_1\neq l_2}^LJ_2(H^{[l_1l_2]}(\mathcal{S}),H^{[l_2l_1]}(\mathcal{S}))=J(\mathcal{H}(\mathcal{S})).
\end{eqnarray}
Here we used the conditions in Theorem 2 for the inequality, and the relationship that 
$$
\sum_{kh}H^{[l]}_{kh}(\mathcal{R})=\sum_{kh}\sum_{ab}P^{[l]}_{ab}R^{[l]}_{ka}(\mathcal{E})R^{[l]}_{hb}(\mathcal{E})=\sum_{ab}P^{[l]}_{ab}\pi^{[l]}_a\pi^{[l]}_b=\sum_{kh}H^{[l]}_{kh}(\mathcal{S})
$$ 
and 
$$
\sum_{kh}H^{[l_1l_2]}_{kh}(\mathcal{R})=\sum_{kh}\sum_{ab}P^{[l_1l_2]}_{ab}R^{[l_1]}_{ka}(\mathcal{E})R^{[l_2]}_{hb}(\mathcal{E})=\sum_{ab}P^{[l_1l_2]}_{ab}\pi^{[l_1]}_a\pi^{[l_2]}_b=\sum_{kh}H^{[l_1l_2]}_{kh}(\mathcal{S}).
$$
We have shown that $\mathcal{S}$ is a maximizer of $J(\mathcal{H}(\mathcal{R}))$.

Next we need to show that $\mathcal{S}$ is the unique maximizer of $J(\mathcal{H}(\mathcal{R}))$.
This can be shown using Lemma 3.2 in Bickel and Chen (2009). 
Since the inequality $J(\mathcal{H}(\mathcal{R}))\le J(\mathcal{H}(\mathcal{S}))$ holds only if $\bigtriangleup_{kh}=\bigtriangleup_{ab}$ whenever $R^{[l]}_{ka}(\mathcal{E})R^{[l]}_{hb}(\mathcal{E})>0$, $l=1,\ldots,L$, and $\bigtriangleup$ does not have two identical columns, using the results in Lemma 3.2, we have $\mathcal{S}$ uniquely maximizes $J(\mathcal{H}(\mathcal{R}))$.

Now that we have shown that $J\left(\mathcal{H}(\mathcal{R})\right)$ is uniquely maximized by $\mathcal{S}$. By the continuity of $J(.)$ in the neighborhood of $\mathcal{S}$, there exists $\delta_n\rightarrow\infty$, such that
$$
J\left(\mathcal{H}(\mathcal{R})\right)-J\left(\mathcal{H}(\mathcal{S})\right)\ge2\epsilon_n\quad\text{for}\quad\eta(\mathcal{E},\mathcal{C})\ge\delta_n.
$$
Here we used the fact that 
\begin{eqnarray*}
\eta(\mathcal{R}(\mathcal{E}),\mathcal{S})&=&\sum_{l=1}^L\sum_{ab}|\pi^{[l]}_bV^{[l]}_{ab}(\mathcal{E})-\pi^{[l]}_bV^{[l]}_{ab}(\mathcal{C})|\\
&\ge&(\min_{l,b}\pi^{[l]}_b)\times\sum_{l=1}^L\sum_{ab}|V_{ab}^{[l]}(\mathcal{E})-V_{ab}^{[l]}(\mathcal{C})|=(\min_{l,b}\pi^{[l]}_b)\times\eta(\mathcal{E},\mathcal{C}).
\end{eqnarray*}

Thus, with (\ref{pop}), we have that
\begin{eqnarray*}
&&P\left(\max_{\mathcal{E}:\ \eta(\mathcal{E},\mathcal{C})\ge\delta_n}J\left(\frac{\mathcal{O}(\mathcal{E})}{\mu_n}\right)\le J\left(\frac{\mathcal{O}(\mathcal{C})}{\mu_n}\right)\right)\\
&&\ge P\left(\left|\max_{\mathcal{E}:\ \eta(\mathcal{E},\mathcal{C})\ge\delta_n}J\left(\frac{\mathcal{O}(\mathcal{E})}{\mu_n}\right)-\max_{\mathcal{E}:\ \eta(\mathcal{E},\mathcal{C})\ge\delta_n}J(\mathcal{H}(\mathcal{R}))\right|<\epsilon_n, \left|J\left(\frac{\mathcal{O}(\mathcal{C})}{\mu_n}\right)-J(\mathcal{H}(\mathcal{S}))\right|\le\epsilon_n\right)\rightarrow1.
\end{eqnarray*}
This implies that 
$$
P(\eta(\mathcal{\hat C},\mathcal{C})\le\delta_n)\rightarrow1,
$$
where 
$$
\mathcal{\hat C}=\arg\max_{\mathcal{E}}J\left(\frac{\mathcal{O}(\mathcal{E})}{\mu_n}\right)
$$
Since
\begin{eqnarray*}
\frac{1}{n}\sum_{l=1}^L\sum_{i=1}^{n_l}I(\hat c_i^{[l]}\neq c_i^{[l]})=\sum_{l=1}^L\sum_k\pi^{[l]}_k(1-V^{[l]}_{kk}(\mathcal{\hat C}))&\le&\sum_{l}^L\sum_k(1-V^{[l]}_{kk}(\mathcal{\hat C}))\\
&=&\frac{1}{2}\sum_{l=1}^L\left(\sum_k(1-V^{[l]}_{kk}(\mathcal{\hat C}))+\sum_{k\neq h}V^{[l]}_{kh}(\mathcal{\hat C})\right)\\
&=&\eta(\mathcal{\hat C},\mathcal{C})/2,
\end{eqnarray*}
we have thus established the consistency property of $\mathcal{\hat C}$.

\section*{References}
\begin{description}
\setlength{\itemsep}{-2ex}
\item Bernstein, S. N. (1924). On a Modification of Chebyshev's Inequality and of the Error Formula of Laplace. \textit{Annals Science Institute SAV. Ukraine} \textbf{4}, 38-49.
\item Bickel, P., and Chen, A. (2009). A non-parametric view of network models and Newman-Girvan and other modularities. \textit{Proceedings of the National Academy of Sciences} \textbf{106}, 21068-21073.
\item Bollob\'{a}s, B., and McKay, B. D. (1986). The number of matchings in random regular graphs and bipartite graphs. \textit{Journal of Combinatorial Theory, Series B} \textbf{41(1)}, 80-91.
\item Zhao, Y., Levina, E., and Zhu, J. (2012). Consistency of community detection in networks under degree-corrected stochastic block models. \textit{The Annals of Statistics} \textbf{40}, 2266-2292.

\end{description}

\end{document}